\documentclass[journal]{IEEEtran}

\usepackage[OT1]{fontenc} 
\usepackage[numbers,sort&compress]{natbib}
\usepackage[cmex10]{amsmath}
\usepackage{amssymb}
\usepackage{bm}
\usepackage{braket}
\usepackage{graphicx}
\usepackage{color}
\usepackage{subfigure}
\usepackage{tabularx}
\usepackage{arydshln}
\usepackage{mathtools}
\usepackage{multirow}
\usepackage{algorithm,algorithmic}
\usepackage{url}
\usepackage{listings}
\usepackage{comment}
\usepackage{CJKutf8}
\usepackage{amsmath} 
\usepackage{makecell}
\usepackage{hyperref}
\usepackage[absolute,overlay]{textpos}

\def\I{\mathbf{I}}
\def\Y{\mathbf{Y}}

\def\H{\mathbf{H}}

\def\z{\mathbf{z}}
\def\a{\mathbf{a}}
\def\w{\mathbf{w}}

\def\V{\mathbf{V}}

\def\W{\mathbf{W}}
\def\X{\mathbf{X}}

\def\b{\mathbf{b}}

\def\dc{d_{\mathrm{c}}}
\def\rep{\mathrm{rep}}
\def\dcmin{d_{\mathrm{c,min}}}
\def\dEmin{d_{\mathrm{E,min}}}
\def\dE{d_{\mathrm{E}}}
\def\dv{d_{\mathrm{v}}}
\def\dh{d_{\mathrm{h}}}
\def\dd{d_{\mathrm{d}}}

\def\bv{\mathbf{v}}
\def\r{\mathbf{r}}
\def\y{\mathbf{y}}
\def\x{\mathbf{x}}

\def\t{\mathbf{t}}
\def\g{\mathbf{g}}
\def\herm{\mathrm{H}}

\usepackage{amsthm}
\newtheorem{lemma}{Lemma}
\newtheorem{theorem}{Theorem}
\newtheorem{proposition}[theorem]{Proposition}

\makeatletter
\def\ps@IEEEtitlepagestyle{%
  \def\@oddfoot{\mycopyrightnotice}%
  \def\@oddhead{\hbox{}\@IEEEheaderstyle\leftmark\hfil\thepage}\relax
  \def\@evenhead{\@IEEEheaderstyle\thepage\hfil\leftmark\hbox{}}\relax
  \def\@evenfoot{}%
}
\def\mycopyrightnotice{%
  \begin{minipage}{\textwidth}
  \centering \scriptsize
\textcopyright 2026 IEEE. Personal use of this material is permitted.
  Permission from IEEE must be obtained for all other uses, in any current or future
  media, including reprinting/republishing this material for advertising or promotional
  purposes, creating new collective works, for resale or redistribution to servers or
  lists, or reuse of any copyrighted component of this work in other works.
  DOI: \href{https://doi.org/10.1109/OJCOMS.2026.3727590}{10.1109/OJCOMS.2026.3727590}
  \end{minipage}
}
\makeatother
\begin{document}
\title{Z-Opt: A Near-Optimal Reduced-Complexity Two-Dimensional Grassmannian Constellation}

\author{
    Kotaro~Shigenaga,~\IEEEmembership{Graduate Student Member,~IEEE},
    Hiroki Iimori,~\IEEEmembership{Member, IEEE},
    Yuto Hama,~\IEEEmembership{Member, IEEE},\\
    Chandan Pradhan,~\IEEEmembership{Member, IEEE},
    Szabolcs Malomsoky, and Naoki~Ishikawa,~\IEEEmembership{Senior~Member,~IEEE}.
    \thanks{K.~Shigenaga and N.~Ishikawa are with the Graduate School of Engineering Science, Yokohama National University, 240-8501 Kanagawa, Japan (e-mail: iskw@ieee.org). H.~Iimori, Y.~Hama, C.~Pradhan, and S.~Malomsoky are with Ericsson Research, Ericsson Japan K. K., Yokohama SYMPHOSTAGE West Tower 12F, 5-1-2 Minato Mirai, Yokohama, 220-0012, Japan (e-mail: [hiroki.iimori, yuto.hama, szabolcs.malomsoky]@ericsson.com).}
    \thanks{This work was partially supported by JSPS KAKENHI Grant Number JP26K00944.}
}

\markboth{\today}
{Shigenaga \MakeLowercase{\textit{et al.}}: Z-Opt: A Near-Optimal Reduced-Complexity Two-Dimensional Grassmannian Constellation}
\maketitle
\TPshowboxesfalse
\begin{textblock*}{\textwidth}(45pt,10pt)
\footnotesize
\centering
Accepted for publication in IEEE Open Journal of the Communications Society. This is the author's version which has not been fully edited and content may change prior to final publication. Citation information: DOI 10.1109/OJCOMS.2026.3727590
\end{textblock*}
\begin{abstract}
Grassmannian constellations are known to achieve the capacity of noncoherent communications over Rayleigh fading channels in the high-SNR regime, yet their efficient construction remains challenging.
In this paper, we present two construction methods for Grassmannian constellations of one-dimensional subspaces in a two-dimensional space, termed S-Opt and Z-Opt, and propose two low-complexity detectors.
Both the construction and detection procedures are performed on the unit sphere, known as the Bloch sphere in quantum computing.
By exploiting the proportionality between the chordal distance on the Grassmann manifold and the Euclidean distance on the Bloch sphere, we derive a corresponding theoretical upper bound on the minimum chordal distance based on the Fejes--T\'oth bound.
The S-Opt constellation is constructed from sphere-packing solutions and achieves the theoretical limit corresponding to the considered optimal Bloch-sphere packings.
The S-Opt detector can be applied to arbitrary Grassmannian constellations on $\mathcal{G}(2,1)$, requiring a number of complex multiplications that scales linearly with the number of receive antennas and logarithmically with the constellation size, together with a number of numerical comparisons that scales logarithmically with the constellation size, while yielding the same detection performance as the GLRT detector.
Furthermore, based on the insight obtained through the S-Opt construction, we propose the Z-Opt constellation, which is constructed by stacking regular polygons on the Bloch sphere and whose minimum chordal distance approaches the derived upper bound over the evaluated constellation sizes.
The Z-Opt detector requires a number of complex multiplications that scales linearly with the number of receive antennas and a number of numerical comparisons that scales logarithmically with the square root of the constellation size, while yielding the same detection performance as the GLRT detector for the Z-Opt constellation.
\end{abstract}


\IEEEpeerreviewmaketitle

\section{Introduction}

\IEEEPARstart{H}{igh-mobility} wireless communications, where channel conditions change rapidly, pose significant challenges for conventional coherent multiple-input multiple-output (MIMO) systems that rely on pilot symbols for channel estimation \cite{ai2014challenges}.
As the channel varies rapidly, a larger proportion of pilot symbols is required, reducing spectral efficiency.
For example, with demodulation reference signals specified in 5G NR, the pilot overhead can reach approximately 30\% depending on the environment \cite{3gpp38211_2018}.
Noncoherent communication, which eliminates the need for channel estimation, has been extensively studied as a promising alternative for such high-mobility scenarios \cite{gohary2019noncoherent,ngo2025noncoherent}.
Such noncoherent signaling methods have also been shown to serve as data-carrying reference signals, enabling simultaneous data transmission and channel estimation for coherent users, thereby improving spectral efficiency in high-mobility scenarios \cite{yu2007informationbearing,endo2024boosting,kato2025maximizing}.

Among approaches to noncoherent communication, including unitary space-time modulation \cite{hochwald2006unitary} and differential space-time coding \cite{hochwald2000differential,tarokh2000differential,ishikawa2018differential}, Grassmannian signaling \cite{zheng2002communication} is of particular interest.
Each codeword spans a distinct subspace that is invariant under multiplication by the channel matrix, enabling communication without channel estimation.
Zheng and Tse \cite{zheng2002communication} showed that the high-SNR capacity of a noncoherent block Rayleigh fading channel, where the channel matrix remains constant over $T$ orthogonal time-frequency resources, can be interpreted as a sphere-packing problem on the Grassmann manifold.
The design of a codebook, known as a Grassmannian constellation, thus reduces to maximizing the minimum chordal distance among the codewords, which is a key performance metric.
For detection, a generalized likelihood ratio test (GLRT) detector is used \cite{warrier2002spectrally} and it is equivalent to a nearest-neighbor search based on the chordal distance.

Conventional construction methods for Grassmannian constellations can be broadly classified into numerical optimization and structured construction.
The numerical optimization approach is referred to as manifold optimization (Man-Opt) \cite{gohary2009noncoherent,ngo2020cubesplit,cuevas2023union,ngo2020multiuser} in this paper.
Man-Opt directly maximizes the minimum chordal distance on the Grassmann manifold, and can thus construct an optimal constellation in terms of the minimum chordal distance.
By contrast, the resulting constellation does not possess any particular structure, and detection requires exhaustive search based on the GLRT detector.
Representative structured construction methods include a method using exponential map (Exp-Map)~\cite{kammoun2003new,kammoun2007noncoherent}, Cube-Split~\cite{ngo2020cubesplit}, which partitions the manifold into cells, and Grass-Lattice~\cite{cuevas2024constellations}, which applies a measure-preserving map from the unit hypercube to the manifold.
These methods offer low-complexity construction and detection by exploiting their inherent structure.
However, the minimum chordal distance and detection performance do not yet match those achieved by Man-Opt with the GLRT detector, leaving room for further improvement.

Several studies in the wireless communications literature have exploited concepts from quantum computing.
For example, stabilizer codes from quantum error correction have been employed as space-time unitary codewords \cite{lanham2019noncoherent,cuvelier2021quantum}, mutually unbiased bases have been utilized for precoder codebook design \cite{inoue2009kerdock} on the Grassmann manifold, and Grassmannian packings based on the Clifford group have been proposed \cite{shor2002family}.
These common connections are summarized in \cite{ishikawa2025quantumaccelerated}.
It is pointed out in \cite{ishikawa2025quantumaccelerated} that the complex Grassmann manifold $\mathcal{G}(2,1)$, which consists of one-dimensional subspaces in a two-dimensional space, can be represented on the surface of a unit sphere, known as the Bloch sphere in quantum computing.

In this paper, we use the Bloch sphere representation of the Grassmann manifold $\mathcal{G}(2,1)$ and construct Grassmannian constellations for $T=2$ orthogonal time-frequency resources and $M=1$ transmit antenna, which corresponds to one-dimensional subspaces in a two-dimensional space\footnote{A reference signal sequence designed for a single-antenna system can be extended to a multi-antenna configuration by assigning orthogonal resources across antennas via code-, time-, or frequency-division multiplexing, thereby ensuring separability of the reference signals.}.
The contributions of this paper are summarized as follows:
\begin{enumerate}
    \item By exploiting the proportionality between the chordal distance on $\mathcal{G}(2,1)$ and the Euclidean distance on the Bloch sphere, we point out that the construction of an optimal Grassmannian constellation on $\mathcal{G}(2,1)$ reduces to the classical unit-sphere packing
problem, which is referred to as the \textit{S-Opt} constellation.
    We also propose the S-Opt detector, which detects an arbitrary Grassmannian constellation on $\mathcal{G}(2,1)$ by performing nearest-neighbor detection on the Bloch sphere.
    The detector achieves the same detection performance as the GLRT detector with a time complexity of $O(N+\log_{2}C)$, where $N$ is the number of receive antennas and $C$ is the constellation size.
    \item Based on the insight obtained from the S-Opt constellation, we propose the \textit{Z-Opt} constellation, which is constructed by stacking regular polygons on the Bloch sphere and approaches the derived upper bound with only $O(\sqrt{C})$ optimization variables.
    Additionally, we propose the Z-Opt detector for the constellation, which achieves the same detection performance as the GLRT detector with a time complexity of $O(N)$, independent of $C$.
    \item We also derive a theoretical upper bound on the minimum chordal distance based on the Fejes--T\'oth bound by exploiting the proportionality between the chordal distance on the Grassmann manifold and the Euclidean distance on the Bloch sphere.
\end{enumerate}

The remainder of this paper is organized as follows.
In Section \ref{sec:system_model}, we describe the system model, the performance metrics for Grassmannian constellations, and the GLRT detector.
In Section \ref{sec:conventional_grassmann_constellation}, we review conventional Grassmannian constellations.
In Sections \ref{sec:s-opt_constellation} and \ref{sec:z-opt_constellation}, we present the S-Opt and Z-Opt constellations, along with their low-complexity detectors.
In Section \ref{sec:performance_results}, we evaluate the performance, and in Section \ref{sec:conclusion}, we conclude this paper.

\section{System Model}
\label{sec:system_model}

In this paper, we consider a block Rayleigh fading channel in which channel coefficients remain constant over $T$ orthogonal time-frequency resources.
Letting $M$ denote the number of transmit antennas and $N$ the number of receive antennas, the received signal is represented as
\begin{align}
    \label{eq:system_model}
    \Y=\sqrt{\frac{T}{M}}\X\H+\W,
\end{align}
where $\X\in\mathbb{C}^{T\times M}$ denotes a space-time codeword, $\H\in\mathbb{C}^{M\times N}$ denotes a channel matrix, and $\W\in\mathbb{C}^{T\times N}$ denotes an additive noise matrix.

When discrete points are selected from the Grassmann manifold $\mathcal{G}\left(T,M\right)$ and used as transmit codewords, they are referred to as Grassmannian codewords.
The set of Grassmannian codewords is referred to as a Grassmannian constellation.
The chordal distance $\dc$ between $\X_{i}$ and $\X_{j}$, which is commonly used as a distance metric between Grassmannian codewords, is expressed as
\begin{align}
    \dc\left(\X_{i},\X_{j}\right)=
    \sqrt{
    M-
    \lVert \X_{i}^{\herm}\X_{j} \rVert_{\mathrm{F}}^{2}
    }.
    \label{eq:chordal_distance}
\end{align}
In particular, when $M=1$, that is, when the two codewords are represented by vectors $\x_{i}$ and $\x_{j}$, the chordal distance is expressed as 
\begin{align}
    \dc\left(\x_{i},\x_{j}\right)=\sqrt{1-\left|\x_{i}^{\herm}\x_{j}\right|^{2}}.
\end{align}
The minimum chordal distance $\dcmin$, which is defined as the smallest chordal distance among all pairs of codewords, is commonly used as a performance metric for constellations and is expressed as
\begin{align}
    \dcmin=\min
    _{1\leq i<j\leq\left| \mathcal{X} \right|}
    {\dc\left(\X_{i},\X_{j}\right)}.
\end{align}
Here, $\mathcal{X}$ denotes a constellation, and $\left| \mathcal{X} \right|$ represents the number of codewords in the constellation, which is denoted by the constellation size $C$.
A larger minimum chordal distance leads to better constellation performance.

In noncoherent detection, the optimal detector for Grassmannian codewords is given by the GLRT detector
\begin{align}
    \hat{\X}=
    \operatorname*{argmax}_{\X\in\mathcal{X}}{
        \lVert \Y^{\herm}\X \rVert_{\mathrm{F}}^{2}
    }.
    \label{eq:GLRT}
\end{align}
By comparing \eqref{eq:chordal_distance} and \eqref{eq:GLRT}, it can be seen that the GLRT detector corresponds to a nearest-neighbor detection in terms of chordal distance.

\section{Conventional Grassmannian Constellations}
\label{sec:conventional_grassmann_constellation}

We review conventional construction methods for Grassmannian constellations.
We introduce Man-Opt \cite{gohary2009noncoherent,ngo2020cubesplit,cuevas2023union,ngo2020multiuser}, a numerical optimization approach, as well as Exp-Map \cite{kammoun2003new,kammoun2007noncoherent}, Cube-Split \cite{ngo2020cubesplit}, and Grass-Lattice \cite{cuevas2024constellations}, which are representative structured construction methods.

\subsection{Man-Opt \cite{gohary2009noncoherent,ngo2020cubesplit,cuevas2023union,ngo2020multiuser}}
Man-Opt is a constellation designed by maximizing the minimum chordal distance between Grassmannian codewords through numerical optimization.
The problem of maximizing the minimum chordal distance is reformulated into a smoothed and differentiable form, which can be expressed as
\begin{align}
    \operatorname*{minimize}_{\mathcal{X}}{
    \log{
    \sum_{1\leq i<j\leq \left| \mathcal{X} \right|}
    {
    \exp{
    \left( \frac{\lVert \X_{i}^{\herm}\X_{j} \rVert_{\mathrm{F}}}{\epsilon} \right)
    }}}
    },
\end{align}
where $\epsilon$ is a smoothing parameter.
Furthermore, when $M=1$, the optimization can be simplified as
\begin{align}
    \operatorname*{minimize}_{\mathcal{X}}{
    \log{
    \sum_{1\leq i<j\leq \left| \mathcal{X} \right|}
    {
    \exp{
    \left( \frac{\left|\x_{i}^{\herm}\x_{j}\right|}{\epsilon} \right)
    }}}
    }.
    \label{eq:Man-Opt_logsumexp}
\end{align}
This optimization can be implemented using Manopt \cite{boumal2014manopt}, a MATLAB toolbox for optimization on manifolds, or its Python counterpart, Pymanopt \cite{townsend2016pymanopt}.

\subsection{Exp-Map \cite{kammoun2003new,kammoun2007noncoherent}}
Exp-Map is a constellation constructed by applying an exponential mapping to complex symbols. 
An arbitrary codeword on $\mathcal{G}(T,M)$ can be expressed as
\begin{align}
    \X=
    \begin{bmatrix}
        \exp{
            \begin{pmatrix}
                \mathbf{0} & \V \\
                -\V^{\herm} & \mathbf{0}
            \end{pmatrix}
        }
    \end{bmatrix}
    \I_{T,M}.
    \label{eq:arbitrary_grassmann_point}
\end{align}
Here, $\V\in\mathbb{C}^{M\times(T-M)}$, $\I_{T,M}=(\I_{M\times M}\; \mathbf{0}_{M\times(T-M)})^{\mathrm{T}}$.
In particular, when $M=1$, \eqref{eq:arbitrary_grassmann_point} can be expressed as
\begin{align}
    \x=
    \begin{pmatrix}
        \cos{\rho}\\
        -\frac{\sin{\rho}}{\rho}\bv
    \end{pmatrix}.
    \label{eq:Exp-Map_for_M=1}
\end{align}
Here, $\bv\in\mathbb{C}^{T-1}$, $\rho=\lVert \bv\rVert$.
When $\rho<\pi/2$, \eqref{eq:Exp-Map_for_M=1} becomes invertible.
The codewords are constructed by treating $\bv$ as $T-1$ complex symbols.
In \cite{kammoun2003new}, PSK symbols are adopted as the complex symbols,
while in \cite{kammoun2007noncoherent}, QAM symbols are used instead.
For example, when $n$ complex symbols are used in the construction, the constellation size becomes $n^{T-1}$.

\subsection{Cube-Split \cite{ngo2020cubesplit}}
Cube-Split is a constellation constructed by partitioning the Grassmann manifold into a collection of bent hypercubes and defining an appropriate mapping onto each of them, so that the resulting codewords are approximately uniformly distributed over the manifold.
This method is applicable only to the case of $M=1$.

First, the Grassmann manifold is partitioned into $T$ cells. $T$ points on the Grassmann manifold are selected as the canonical basis vectors $\{ \mathbf{e}_{1},\cdots,\mathbf{e}_{T} \}$, and the cells are defined as the Voronoi regions with respect to the chordal distance induced by these vectors.

Next, a grid is constructed in the Euclidean space.
A Grassmannian codeword has $T-1$ complex dimensions, which corresponds to $2(T-1)$ real dimensions.
When $B_{j}$ bits are used to represent each real dimension within the interval $\left(0,1\right)$, the corresponding set $A_{j}$ can be expressed as
\begin{align}
    A_{j}=
    \left\{
    \frac{1}{2^{B_{j}+1}},
    \frac{3}{2^{B_{j}+1}},
    \cdots,
    \frac{2^{B_{j}+1}-1}{2^{B_{j}+1}}
    \right\},
\end{align}
where $1\leq j\leq 2(T-1)$.
Compute the Cartesian product of all the sets $A_{j}$ defined above, $\bigotimes_{j=1}^{2(T-1)}A_{j}$, to form the grid.
Denote a grid by $\a\in\bigotimes_{j=1}^{2(T-1)}A_{j}$.

Finally, map $\a$ onto a Grassmann manifold.
Using the mapping $\xi_{T-1}:\a\mapsto\t=[\t_{1},\cdots,\t_{T-1}]^{\mathrm{T}}$, the mapping onto the $i$-th cell is defined as
\begin{align}
    \g_{i}\left(\a\right)=
    \frac{1}{\sqrt{1+\sum_{j=1}^{T-1}{\left|t_{j}\right|^{2}}}}
    \left[ t_{1}\;\cdots\;t_{i-1}\;1\;t_{i}\cdots\;t_{T-1} \right]^{\mathrm{T}},
\end{align}
where $1\leq i\leq T$.
For $T=2$, the mapping $\xi_{1}$ is defined as
\begin{align}
    \xi_{1}\left(\a\right)\triangleq
    \sqrt{
    \frac{
    1-\exp{
    \left( -\frac{\left|w\right|^{2}}{2} \right)
    }
    }{
    1+\exp{
    \left( -\frac{\left|w\right|^{2}}{2} \right)
    }
    }
    }
    \frac{w}{\left|w\right|}
    \label{eq:xi_1}
\end{align}
where $w=\mathcal{N}^{-1}(a_{1})+j\mathcal{N}^{-1}(a_{2})$.
For $T>2$, using \eqref{eq:xi_1}, the mapping $\xi_{T-1}$ is given by
\begin{align}
    \xi_{T-1}\left(\a\right)=
    \left[
    \xi_{1}\left(\left[ a_{1}\;a_{2} \right]^{\mathrm{T}}\right)
    \;\cdots\;
    \xi_{1}\left(\left[ a_{2T-3}\;a_{2T-2} \right]^{\mathrm{T}}\right)
    \right]^{\mathrm{T}}.
\end{align}

\subsection{Grass-Lattice \cite{cuevas2024constellations}}
Grass-Lattice is a constellation constructed by mapping points from the unit hypercube onto the Grassmann manifold through a measure-preserving mapping, aiming to achieve a uniform distribution of codewords over the manifold.
The mapping is composed of three stages, $\vartheta=\vartheta_{3}\circ\vartheta_{2}\circ\vartheta_{1}$, and can be expressed as
\begin{align}
    \vartheta:\mathcal{I}=\left(0,1\right)\times \cdots\times\left(0,1\right)\;\rightarrow\;\mathcal{G}\left(T,1\right).
\end{align}
Elements in $\mathcal{I}$ are denoted by
\begin{align}
    \left( \a,\b \right)=
    \left(
    a_{1},\dots,a_{T-1},b_{1},\dots ,b_{T-1}
    \right)
    ,\;
    a_{k},b_{k}\in\left(0,1\right).
\end{align}
To make each element's interval closed instead of open, we set the lowest point to $\alpha>0$ and the highest point to $1-\alpha<1$, resulting in the interval $[\alpha,1-\alpha]$.
In this case, if $B_r$ denotes the number of bits assigned to each element, the corresponding coordinates are expressed as 
\begin{align}
    \hat{x}_{p}=\alpha+p\frac{1-2\alpha}{2^{B_{r}}-1},\;0\leq p\leq 2^{B_{r}}-1.
\end{align}

$\vartheta_{1}$ maps points uniformly distributed in the unit hypercube to points normally distributed in $\mathbb{C}^{T-1}$.
Compute
\begin{align}
    z_{k}=F^{-1}\left(a_{k}\right)+jF^{-1}\left( b_{k} \right),\;
    k=1,\dots,T-1,
\end{align}
where $F(x)$ is the cdf of a $\mathcal{N}(0,1/2)$.
The point $\mathbf{z}$ is isotropically distributed as $\mathbf{z}\sim\mathcal{CN}(\mathbf{0},\I_{T-1})$.

$\vartheta_{2}$ maps normally distributed points in $\mathbb{C}^{T-1}$ to points uniformly distributed in the unit ball
\begin{align}
    \mathbb{B}_{\mathbb{C}^{T-1}}\left(0,1\right)=
    \left\{
    \mathbf{w}\in\mathbb{C}^{T-1},\;\lVert \mathbf{w} \rVert <1
    \right\}.
\end{align}
Compute $\mathbf{w}=\mathbf{z}f_{T-1}(\lVert \mathbf{z} \rVert)$, where $f_{T-1}(\cdot)$ is given by
\begin{align}
    f_{T-1}(t)=
    \frac{1}{t}
    \left(
    1-e^{-t^{2}}
    \sum_{k=0}^{T-2}{
    \frac{t^{2k}}{k!}
    }
    \right)^{1/\left(2\left(T-1\right)\right)}.
\end{align}

Finally, $\vartheta_{3}$ maps uniformly distributed points in the unit ball to codewords uniformly distributed in $\mathcal{G}(T,1)$.
Compute
\begin{align}
    \x=
    \begin{pmatrix}
        \sqrt{1-\lVert \w \rVert^{2}}\\
        \w
    \end{pmatrix}.
\end{align}
The resulting constellation has a total size of $2^{2\left(T-1\right)B_{r}}$.

\section{S-Opt Constellation}
\label{sec:s-opt_constellation}

First, we present a Grassmannian constellation, termed S-Opt constellation, for the case of $M=1$ and $T=2$, along with its low-complexity detector.
Both the construction and detection procedures are performed on the Bloch sphere in quantum computing.
The name S-Opt derives from the use of solutions to the unit-sphere packing problem.
Since many exact or near-optimal solutions are already known, no optimization is required for the construction.
The time complexity of the low-complexity detector is $O(N+\log_{2}{C})$, and it is applicable to any Grassmannian constellation on $\mathcal{G}(2,1)$.

\subsection{Construction Method}
The outline of the S-Opt construction procedure is as follows:
\begin{enumerate}
    \item Generate a set of points $\r_{i}\in\mathbb{R}^{3}\ (1\leq i\leq C)$ that are distributed on the Bloch sphere with respect to the Euclidean distance.
    \item For each point $\r_{i}$, compute its polar angle $\theta_{i}$ and azimuth angle $\phi_{i}$ in spherical coordinates.
    \item Compute $\x_{i}$ from $\theta_{i}$ and $\phi_{i}$.
\end{enumerate}

First, we prepare the coordinates of optimally or sub-optimally arranged points on the Bloch sphere, denoted by $\r_{i}\ (1\leq i\leq C)$.
Each $\r_{i}$ can be expressed using two parameters, $\theta_{i}\ (0\leq \theta_{i}\leq \pi)$ and $\phi_{i}\ ( 0\leq\phi_{i}\leq 2\pi)$ as
\begin{align}
    \r_{i}\left( \theta_{i},\phi_{i} \right)=
    \begin{pmatrix}
        r_{i,x}\\
        r_{i,y}\\
        r_{i,z}
    \end{pmatrix}
    =
    \begin{pmatrix}
        \sin{\theta_{i}}\cos{\phi_{i}}\\
        \sin{\theta_{i}}\sin{\phi_{i}}\\
        \cos{\theta_{i}}
    \end{pmatrix}.
    \label{eq:bloch_vector}
\end{align}
The problem of optimal point configurations on a sphere is a classical problem in mathematics, known as the Tammes problem. Optimal arrangements have been rigorously proven for up to 14 points and also for 24 points \cite{musin2015tammes}. For larger numbers of points, near-optimal numerical solutions are available.
For example, the library \cite{cohn2024table} provides them for up to 1024 points.

From the coordinates $\r_{i}$, the parameters $\theta_{i}$ and $\phi_{i}$ can be computed as
\begin{align}
    \theta_{i}&=\arccos{r_{i,z}}\;\text{and}\\
    \phi_{i}&=\arg{\left( r_{i,x}+jr_{i,y} \right)}.
\end{align}

Next, we construct the Grassmannian codeword from $\theta_{i}$ and $\phi_{i}$ as
\begin{align}
    \x_{i}=
    \begin{pmatrix}
        \cos{\frac{\theta_{i}}{2}}\\
        e^{j\phi_{i}}\sin{\frac{\theta_{i}}{2}}
    \end{pmatrix}.
    \label{eq:grassmann_codeward}
\end{align}
With this construction, the Euclidean distance $\dE(\r_{i},\r_{j})$ between two points $\r_{i},\r_{j}$ on the Bloch sphere and the chordal distance $\dc(\x_{i},\x_{j})$ between the corresponding two codewords $\x_{i},\x_{j}$ on the Grassmann manifold satisfy the following relationship:
\begin{align}
    \dE\left(\r_{i},\r_{j}\right)=2\dc\left(\x_{i},\x_{j}\right).
    \label{eq:relation_between_dch_and_deu}
\end{align}
In other words, the chordal distance between codewords on the Grassmann manifold is equal to one-half of the Euclidean distance between the corresponding points on the Bloch sphere.
Therefore, the upper bound of the minimum chordal distance for optimally arranged codewords on the Grassmann manifold can be obtained by taking one-half of the upper bound of the minimum Euclidean distance for optimally arranged points on the Bloch sphere.
For example, the following theoretical upper bound on the minimum Euclidean distance between optimally arranged points on the unit sphere is known \cite{fejes1943über}:
\begin{align}
    \dEmin\leq
    \sqrt{4-\csc^{2}{\left(\frac{\pi C}{6\left(C-2\right)}\right)}}.
\end{align}
By taking one-half of this upper bound, we obtain the following theoretical upper bound on the minimum chordal distance on the Grassmann manifold:
\begin{align}
    \dcmin\leq \frac{1}{2}
    \sqrt{4-\csc^{2}{\left(\frac{\pi C}{6\left(C-2\right)}\right)}}.
    \label{eq:fejes_toth}
\end{align}
Note that the equality is known to hold for $C=3,4,6$, and $12$, but is not generally attained for other values of $C$.

\subsection{Low-Complexity Detector}
\label{subsec:S-Opt_detector}
The received signal is detected by mapping the normalized received point on the Grassmann manifold onto the surface of the Bloch sphere.
The outline of the detection procedure is as follows.
\begin{enumerate}
    \item Normalize the received signal $\y=\left[ y_{1}\;y_{2} \right]^{\mathrm{T}}$ onto the Grassmann manifold.
    \item Compute the polar angle $\theta_{z}$ and azimuth angle $\phi_{z}$ from the normalized received point, and map it onto the Bloch sphere.
    \item Perform nearest-neighbor detection on the Bloch sphere in terms of Euclidean distance, using the KD-tree algorithm.
\end{enumerate}

First, the received signal $\y$ is normalized in terms of both norm and phase.
The norm is set to $1$, and the phase is adjusted by multiplying $\y$ by its conjugate phase so that $y_{1}$ becomes real-valued.
The resulting point on the Grassmann manifold obtained from the normalized $\y$ is expressed as
\begin{align}
    \z=\exp(-j\arg{y_{1}})\frac{\y}{\lVert \y \rVert}.
    \label{eq:normalized_y}
\end{align}
Here, when $N>1$, the received signal is represented as $\Y\in\mathbb{C}^{T\times N}$.
In this case, the singular value decomposition (SVD) of $\Y$ is computed,  and the left singular vector corresponding to the largest singular value is taken as $\y$ and normalized \cite{ngo2020cubesplit}.

Next, the normalized received signal $\z=\left[z_{1}\;z_{2}\right]^{\mathrm{T}}$ is mapped onto the Bloch sphere.
The polar angle $\theta_{z}$ and azimuth angle $\phi_{z}$ on the Bloch sphere are computed as
\begin{align}
    \theta_{z}&=2\arccos{z_{1}}\;\text{and}\\
    \phi_{z}&=\arg{z_{2}}.
    \label{eq:get_theta_z_and_phi_z}
\end{align}
By substituting these values into \eqref{eq:bloch_vector}, the coordinates $\r_{z}(\theta_{z},\phi_{z})$, are obtained.

Finally, nearest-neighbor detection is performed on the Bloch sphere in terms of Euclidean distance.
According to \eqref{eq:relation_between_dch_and_deu} and \eqref{eq:GLRT}, nearest-neighbor detection on the Bloch sphere in terms of Euclidean distance is equivalent to the GLRT detector.
By employing the KD-tree algorithm \cite{pedregosa2011scikitlearn}, the detection can be performed with a time complexity of $O(\log_{2}{C})$ and a space complexity of $O(C)$.

For these reasons, this low-complexity detector can be applied not only to the S-Opt constellation but also to general Grassmannian constellations on $\mathcal{G}(2,1)$ constructed by other methods.

\section{Z-Opt Constellation}
\label{sec:z-opt_constellation}
From observations of the S-Opt constellation, which is a solution to the unit-sphere packing problem, we notice that points on the Bloch sphere can be constructed by layering regular polygons.
Based on this observation, in this section, we propose Z-Opt, a constellation with a specific geometric structure.
The name derives from the fact that only the inter-layer distances, i.e., variables along the Z-axis, need to be optimized.
Specifically, the number of optimization variables and the number of distance metric evaluations per objective function evaluation are both reduced to $O(\sqrt{C})$, while the minimum chordal distance approaches the theoretical upper bound.
The structure of the constellation enables low-complexity detection with time complexity $O(N)$ and space complexity $O(\sqrt{C})$.

\subsection{Construction Method}
The outline of the Z-Opt construction procedure is as follows:
\begin{enumerate}
    \item Construct a structure by rotating and stacking regular polygons on the Bloch sphere, as shown in Table~\ref{table:structure_table}.
    \item Create a structure diagram and, by exploiting its symmetry, reduce the number of optimization variables and identify candidate distances that may become the minimum.
    \item Optimize the variables so as to maximize the minimum among these candidate distances.
    \item From the optimized positions of the regular polygons, compute all $\theta$ and $\phi$, and construct the Grassmannian codewords.
\end{enumerate}

\begin{figure}[tb]
	\centering
	\subfigure[
        $B=4$
        \label{subfig:globe_Z-Opt_B=4}
    ]{
		\includegraphics[clip, scale=0.45]{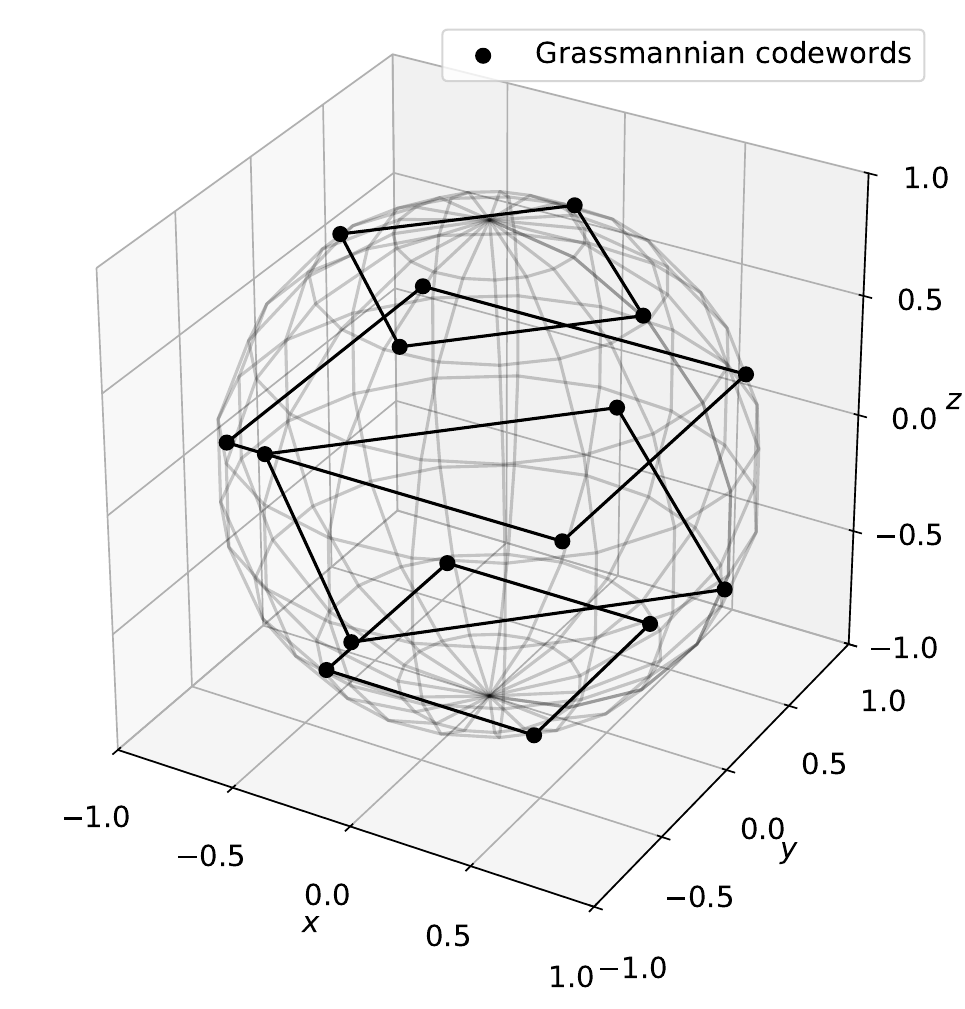}
	}
	\subfigure[
        $B=5$
        \label{subfig:globe_Z-Opt_B=5}
    ]{
		\includegraphics[clip, scale=0.45]{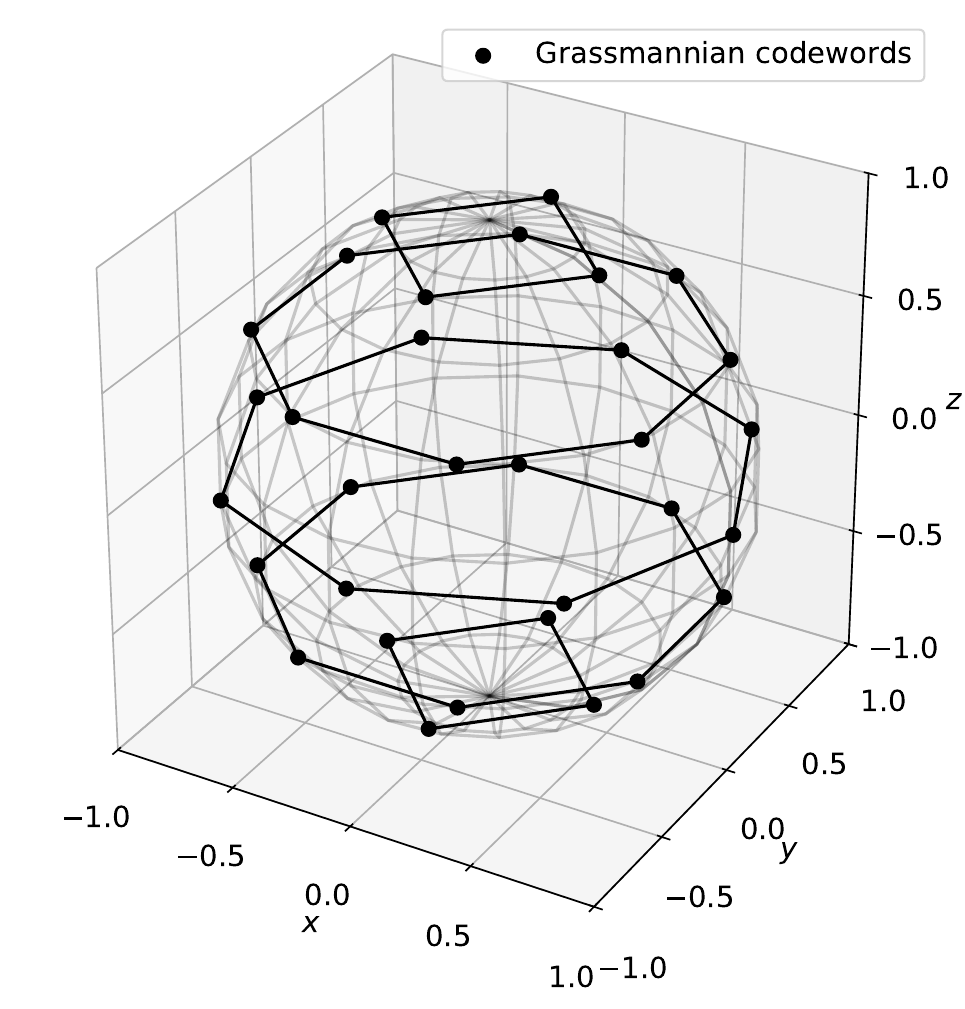}
	}
	\caption{
        Proposed Z-Opt structure of points on the Bloch sphere.
        The points connected by a black line are at the same height.
        \label{fig:globe_Z-Opt}
    }
\end{figure}

We first determine the structural arrangement of points on the Bloch sphere.
As illustrated in Fig.~\ref{fig:globe_Z-Opt}, the Z-Opt constellation is constructed as an $l$-layer structure on the Bloch sphere, where rotated regular polygons are stacked at different heights.
The definition of this structure is summarized in Table~\ref{table:structure_table}.
\begin{table}[tb]
    \begin{center}
    \renewcommand{\arraystretch}{1.0}
    \caption{Structural parameters of Z-Opt \\ for the information bits $B$}
    \label{table:structure_table}
    \begin{tabular}{|r|r|r|c|r|r|}
        \hline
        $B$ & $C$ & $l$ & $Z_{l}=\left( z_{1},\cdots,z_{l} \right)$ & $z_{\mathrm{max}}$ & $n_{v}$ \\
        \hline
         1 &      2 &   1 & $(2)$ &   2 &   1 \\
         \hline
         2 &      4 &   2 & $(2,2)$ &   2 &   1 \\
         \hline
         3 &      8 &   2 & $(4,4)$ &   4 &   1 \\
         \hline
         4 &     16 &   4 & $(4, 4, 4, 4) = \rep(4,4)$ &   4 &   2 \\
         \hline
         5 &     32 &   5 & $(4,\rep(8,3),4)$ &   8 &   2 \\
         \hline
         6 &     64 &   8 & $\rep(8,8)$   &   8 &   4 \\
         \hline
         7 &    128 &   9 & $(8,\rep(16,7),8)$ &  16 &   4 \\
         \hline
         8 &    256 &  16 & $\rep(16,16)$ &  16 &   8 \\
         \hline
         9 &    512 &  32 & $\rep(16,32)$ &  16 &  16 \\
        \hline
        10 &  1,024 &  32 & $\rep(32,32)$ &  32 &  16 \\
        \hline
        11 &  2,048 &  64 & $\rep(32,64)$ &  32 &  32 \\
        \hline
        12 &  4,096 &  64 & $\rep(64,64)$ &  64 &  32 \\
        \hline
        13 &  8,192 & 128 & $\rep(64,128)$ &  64 &  64 \\
        \hline
        14 & 16,384 & 128 & $\rep(128,128)$ & 128 &  64 \\
        \hline
        15 & 32,768 & 256 & $\rep(128,256)$ & 128 & 128 \\
        \hline
        16 & 65,536 & 256 & $\rep(256,256)$ & 256 & 128 \\
        \hline
    \end{tabular}
    \end{center}
\end{table}
Here, $B$ denotes the number of information bits, and $C\ (=2^{B})$ is the number of codewords.
The parameter $l$ represents the total number of polygons, and the sequence $Z_{l}$ denotes the number of points $z_{n}\ (1 \leq n \leq l)$ contained in the $n$-th polygon counted from the top of the Bloch sphere, where $C = z_{1} + \cdots + z_{l}$.
Here, $\rep(a,b)$ indicates that $a$ is repeated $b$ times.
For example, $\rep(4,4) = (4,4,4,4)$ and $(4,\rep(8,3),4) = (4,8,8,8,4)$.
The parameter $z_{\mathrm{max}}$ denotes the maximum value in the sequence $Z_{l}$, i.e., $z_{\mathrm{max}} = \max (z_{1}, \cdots, z_{l})$.
Furthermore, $n_{v}\ (=\lfloor l/2 \rfloor)$ represents the number of variables to be optimized. 
At the $n$-th height from the top, points are arranged to form a regular $z_{n}$-gon.
To account for the distances between points at adjacent heights, the regular polygons are stacked while being rotated.
For $B\notin\{5,7\}$, the regular $z_{n}$-gon is stacked with a rotation of $\pi/z_{\mathrm{max}}$.
For $B\in\{5,7\}$, $l$ is odd, and the first and the $l$-th layers contain half as many points, $z_{\mathrm{max}}/2$, as other layers.
The polygons at these layers are initially regarded as regular $z_{\mathrm{max}}$-gons, from which every other point is removed to obtain the intended structure.
Note that for $B\in\{5,7\}$, it is also possible to construct $Z_l$ (e.g., $Z_l=(\mathrm{rep}(4,8)), (\mathrm{rep}(8,16))$) such that $l$ is even and the number of points in each polygon is identical.
This is not our choice, as this construction results in a slight degradation in the minimum chordal distance.

Next, we consider a structure diagram that represents the points on the Bloch sphere projected onto a plane.
By plotting $\theta$ on the vertical axis and $\phi$ on the horizontal axis, the structure can be represented as shown by the black dots in Fig.~\ref{fig:structure_diagram=B=4,5}.
\begin{figure}[tb]
	\centering
	\subfigure[
        $B=4$
        \label{subfig:structure_diagram_B=4}
    ]{
		\includegraphics[clip, scale=0.42]{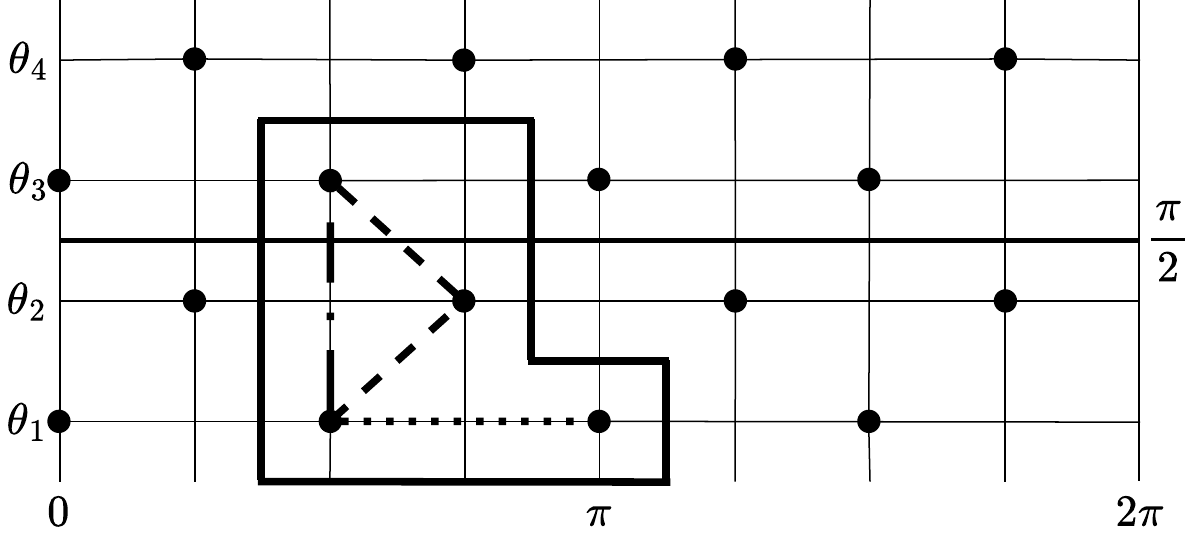}
	}
	\subfigure[
        $B=5$
        \label{subfig:structure_diagram_B=5}
    ]{
		\includegraphics[clip, scale=0.42]{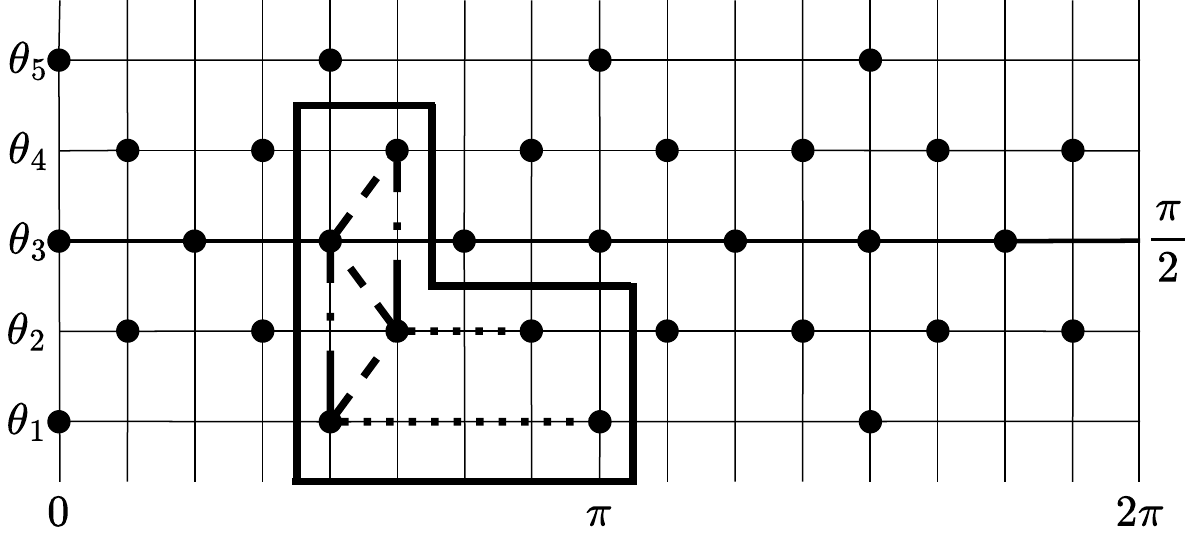}
	}
	\caption{
        Structure diagrams for $B\in\{4,5\}$.
        The black dotted lines indicate candidate distances that may become minimal when optimizing $\theta$. 
        \label{fig:structure_diagram=B=4,5}
    }
\end{figure}
Since $r_{z}=\cos{\theta}$, the $n$-th height from the top of the Bloch sphere corresponds to the height $\theta_{n}$ in the structure diagram. 

\begin{figure}[tb]
	\centering
	\subfigure[
        $B\notin\{5,7\}$
        \label{subfig:structure_diagram_Bnotin57}
    ]{
		\includegraphics[clip, scale=0.42]{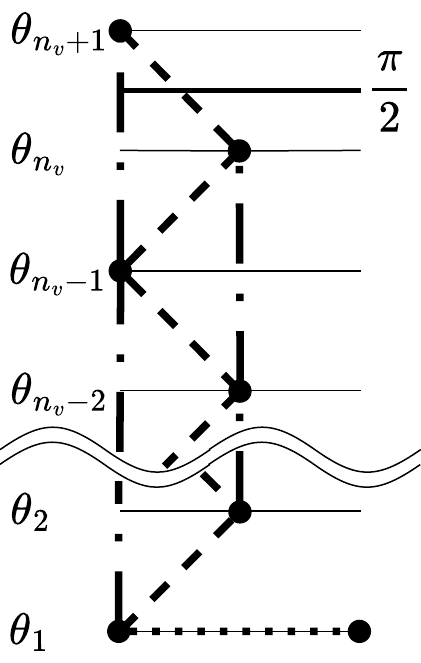}
	}
    \subfigure[
        $B\in\{5,7\}$
        \label{subfig:structure_diagram_Bin57}
    ]{
		\includegraphics[clip, scale=0.42]{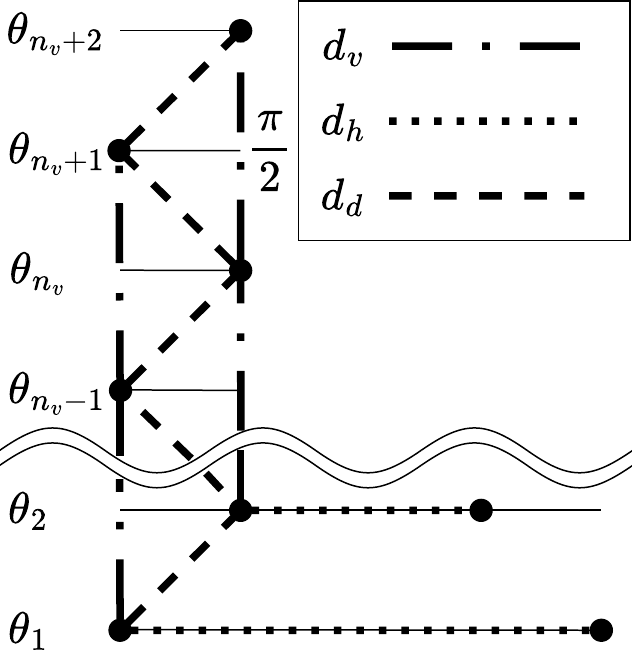}
	}
	\caption{
        Generalization of $\theta$ and the candidate distances that may become the minimum.
        \label{fig:structure_diagram_general}
    }
\end{figure}

Using this structure diagram, we reduce both the number of optimization variables and the number of distance calculations required in the objective function.
As will be described later, for $B\in\{1,2,3\}$, closed-form expressions for $\theta$ can be derived without optimization.
Hence, optimization is not performed in these cases. Therefore, in what follows, we assume $B$ to be an integer in the range $4\leq B\leq 16$.
For $B\notin\{5,7\}$, the parameters $\theta_{i}$ for $1\leq i\leq l$ are assumed to satisfy
\begin{align}
    0<\theta_{1}<\cdots<\theta_{n_{v}}<\frac{\pi}{2}<\theta_{n_{v}+1}<\cdots<\theta_{l}<\pi.
\end{align}
For $B\in\{5,7\}$, they are assumed to satisfy
\begin{align}
    0<\theta_{1}<\cdots<\theta_{n_{v}}<\theta_{n_{v}+1}=\frac{\pi}{2}<\theta_{n_{v}+2}<\cdots<\theta_{l}<\pi.
\end{align}
In both cases, the following relation also holds:
\begin{align}
    \theta_{l-\left(i-1\right)}=\pi-\theta_{i}\ \left(i=1,\cdots,n_{v}\right).
\end{align}
By imposing these constraints, the number of variables to be optimized, denoted as $n_{v}=\lfloor l/2 \rfloor$, can be reduced from the original $l$ parameters.
Moreover, due to symmetry, the candidate distances that may attain the minimal value can be narrowed down.
In Fig.~\ref{fig:structure_diagram=B=4,5}, the distances indicated by dashed lines enclosed by black frames can be generalized as shown in Fig.~\ref{fig:structure_diagram_general}.
Here, the candidate distances that may attain the minimal value can be classified into vertical, horizontal, and diagonal relationships between points.
The functions for computing these distances can be expressed as
\begin{align}
    \dv\left(\theta_{i},\theta_{j}\right)
    &=2\sin{\frac{\left|\theta_{i}-\theta_{j}\right|}{2}},
    \label{eq:dv}\\
    \dh\left(\theta,\Delta\phi\right)
    &=2\sin{\theta}\sin{\frac{\Delta\phi}{2}}\;\text{and}
    \label{eq:dh}\\
    \dd\left(\theta_{i},\theta_{j},\Delta\phi\right)
    &=2\sqrt{\sin^{2}{\frac{\theta_{i}-\theta_{j}}{2}}+\sin{\theta_{i}}\sin{\theta_{j}}\sin^{2}{\frac{\Delta\phi}{2}}}.
    \label{eq:dd}
\end{align}
In the vertical case, where $\phi=\phi_{i}=\phi_{j}$, the distance is expressed as $\dv(\theta_{i},\theta_{j})$.
In the horizontal case, where $\theta=\theta_{i}=\theta_{j}$, we define $\Delta\phi=|\phi_{i}-\phi_{j}|$ and denote the distance as $\dh(\theta,\Delta\phi)$.
In the diagonal case, where both $\theta$ and $\phi$ differ, the distance is denoted by $\dd(\theta_{i},\theta_{j},\Delta\phi)$.
Accordingly, the sets of candidate distances that may become the minimum are given by
\begin{align}
    \mathcal{V}&=\left\{\;\dv\left( \theta_{i},\theta_{i+2} \right)\;\middle|\; 1\leq i\leq n_{v}^{\prime}-1\;\right\},\\
    \mathcal{H}&=\left\{\;\dh\left(\theta_{i},\frac{2\pi}{z_{i}}\right)\;\middle|\; i\in \mathcal{I}\;\right\}\;\text{and}\\
    \mathcal{D}&=\left\{\; \dd\left(\theta_{i},\theta_{i+1},\frac{\pi}{z_{\mathrm{max}}}\right)\;\middle|\;1\leq i\leq n_{v}^{\prime} \;\right\},
\end{align}
where we have
\begin{align}
    n_{v}^{\prime}=
    \begin{cases}
        n_{v},&B\notin\left\{5,7\right\},\\
        n_{v}+1,&B\in\left\{5,7\right\},
    \end{cases}
\end{align}
and
\begin{align}
    \mathcal{I}=
    \begin{cases}
        \left\{ 1 \right\},&B\notin\left\{5,7\right\},\\
        \left\{ 1,2 \right\},&B\in\left\{5,7\right\}.
    \end{cases}
\end{align}
Finally, the set of candidate distances that may attain the minimum is given by
\begin{align}
    \mathcal{V}\cup \mathcal{H}\cup \mathcal{D}.
\end{align}
For $B\notin\{5,7\}$, there is $1$ horizontal line, $n_{v}-1$ vertical lines, and $n_{v}$ diagonal lines, giving a total of $2n_{v}$ candidate distances.
For $B\in\{5,7\}$, there are $2$ horizontal lines, $n_{v}$ vertical lines, and $n_{v}+1$ diagonal lines, resulting in a total of $2n_{v}+3$ candidate distances.

Next, we optimize $\theta_{1},\cdots,\theta_{n_{v}}$ so as to maximize the minimum among the candidate distances.
This optimization problem can be significantly simplified to
\begin{align}
    \operatorname*{maximize}_{\theta_{1},\cdots,\theta_{n_{v}}}{\min{\left(\mathcal{V}\cup \mathcal{H}\cup \mathcal{D}\right)}}.
\end{align}
The optimization problem can be numerically solved using, for example, the dual annealing algorithm in the SciPy optimization library \cite{2020SciPy-NMeth}.
After optimization, we obtain the sequence $\Theta_{n}$, which stores all values of $\theta$, from the optimized $\theta_{1},\cdots,\theta_{n_{v}}$. For $B\notin\{5,7\}$, it is given by
\begin{align}
    \Theta_{n}=
    \begin{cases}
        \theta_{n},&1\leq n\leq n_{v},\\
        \pi-\theta_{l-n+1},&n_{v}+1\leq n\leq l.
    \end{cases}
    \label{eq:Theta_n_Bneq5,7}
\end{align}
For $B\in\{5,7\}$, it is given by
\begin{align}
    \Theta_{n}=
    \begin{cases}
        \theta_{n},& 1\leq n\leq n_{v}, \\
        \frac{\pi}{2},& n=n_{v}+1, \\
        \pi-\theta_{l-n+1},& n_{v}+2\leq n\leq l.
    \end{cases}
    \label{eq:Theta_n_B=5,7}
\end{align}

Finally, we determine the values of $\phi$ for the points included at each height $\theta$ and construct all Grassmannian codewords as
\begin{align}
    \mathcal{X}=
    \left\{
    \X_{C_{m-1}+n}=
    \begin{pmatrix}
        \cos{\frac{\theta_{m}}{2}}\\
        e^{j\phi_{m,n}}\sin{\frac{\theta_{m}}{2}}
    \end{pmatrix}
    \;\middle|\;
    \begin{array}{l}
        1\leq m\leq l,\\
        1\leq n\leq z_{m}.
    \end{array}
    \right\}
    \label{eq:z_opt_constellation}
\end{align}
Here, $C_m$ denotes the cumulative number of codewords up to the $m$-th layer and is defined as
\begin{align}
    C_{m}=
    \begin{cases}
        0, &m=0, \\
        \sum_{k=1}^{m}{z_{k}} ,&m\ge 1,
    \end{cases}
\end{align}
and taking $m$ modulo $2$, $\phi_{m,n}$ is given by
\begin{align}
    \phi_{m,n}=
    \begin{cases}
        \frac{2\left( n-1 \right)\pi}{z_{m}},&m\equiv 1,\\
        \frac{2\left(n-1\right)\pi}{z_{m}}+\frac{\pi}{z_{\mathrm{max}}},&m\equiv 0.
    \end{cases}
    \label{eq:phi_n}
\end{align}

For $B\in\{1,2,3\}$, the optimal configuration is achieved on the Bloch sphere.
These parameters can be expressed in closed form.
The structure diagram is shown in Fig.~\ref{fig:structure_diagram=B=1,2,3}.
\begin{figure}[tb]
	\centering
	\subfigure[
        $B=1$
        \label{subfig:structure_diagram_B=1}
    ]{
		\includegraphics[clip, scale=0.42]{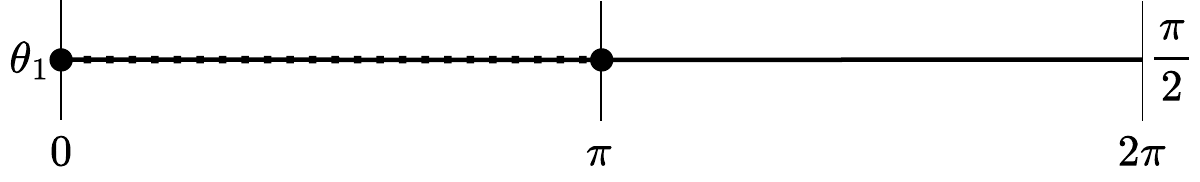}
	}
	\subfigure[
        $B=2$
        \label{subfig:structure_diagram_B=2}
    ]{
		\includegraphics[clip, scale=0.42]{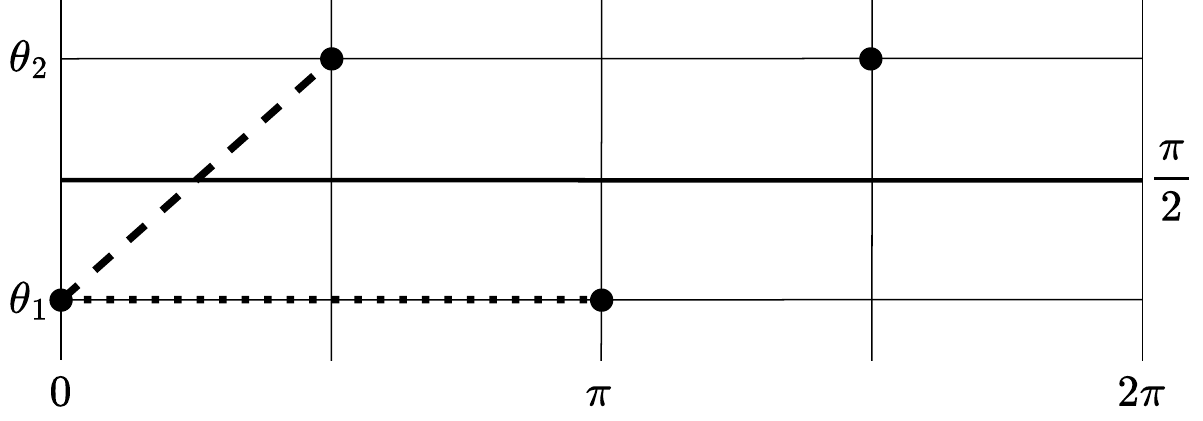}
	}
    \subfigure[
        $B=3$
        \label{subfig:structure_diagram_B=3}
    ]{
		\includegraphics[clip, scale=0.42]{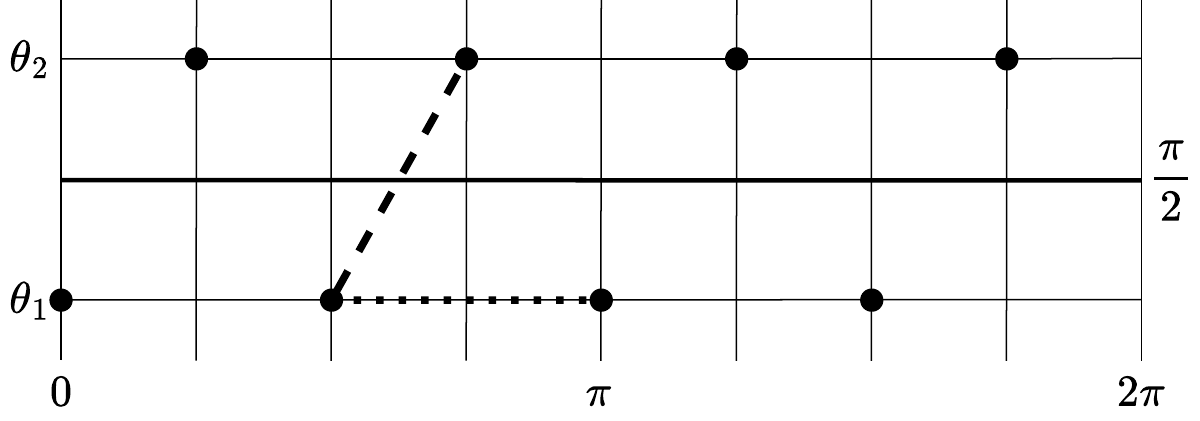}
	}
	\caption{
        Structure diagrams for $B\in\{1,2,3\}$.
        \label{fig:structure_diagram=B=1,2,3}
    }
\end{figure}
For $B=1$, when $\theta_{1}=\pi/2$, the distance between the two points becomes $2$, which is the maximum value.
In this case, the two points are located at diametrically opposite positions on the Bloch sphere.
For $B=2$, by imposing the conditions that two distances indicated by the dotted lines are equal and that $\theta_{2}=\pi-\theta_{1}$, we obtain 
\begin{align}
    \theta_{1}&=\arctan{\sqrt{2}}\,\text{and}\\
    \theta_{2}&=\pi-\arctan{\sqrt{2}}.
\end{align}
In this case, the four points form a regular tetrahedron and the minimum distance is $\frac{2\sqrt{6}}{3}$.
For $B=3$, in a similar manner, we obtain
\begin{align}
    \theta_{1}&=\arctan{\sqrt{2\sqrt{2}}}\;\text{and}\\
    \theta_{2}&=\pi-\arctan{\sqrt{2\sqrt{2}}}.
\end{align}
In this case, the eight points form a square antiprism and the minimum distance is $2\sqrt{\frac{4-\sqrt{2}}{7}}$.

\subsection{Low-Complexity Detector}
\label{subsec:Z-Opt_detector}
\begin{figure}[tb]
	\centering
	\subfigure[
        Visualization of the nearest-neighbor regions.
        \label{subfig:domain_B=6}
    ]{
		\includegraphics[clip, scale=0.37]{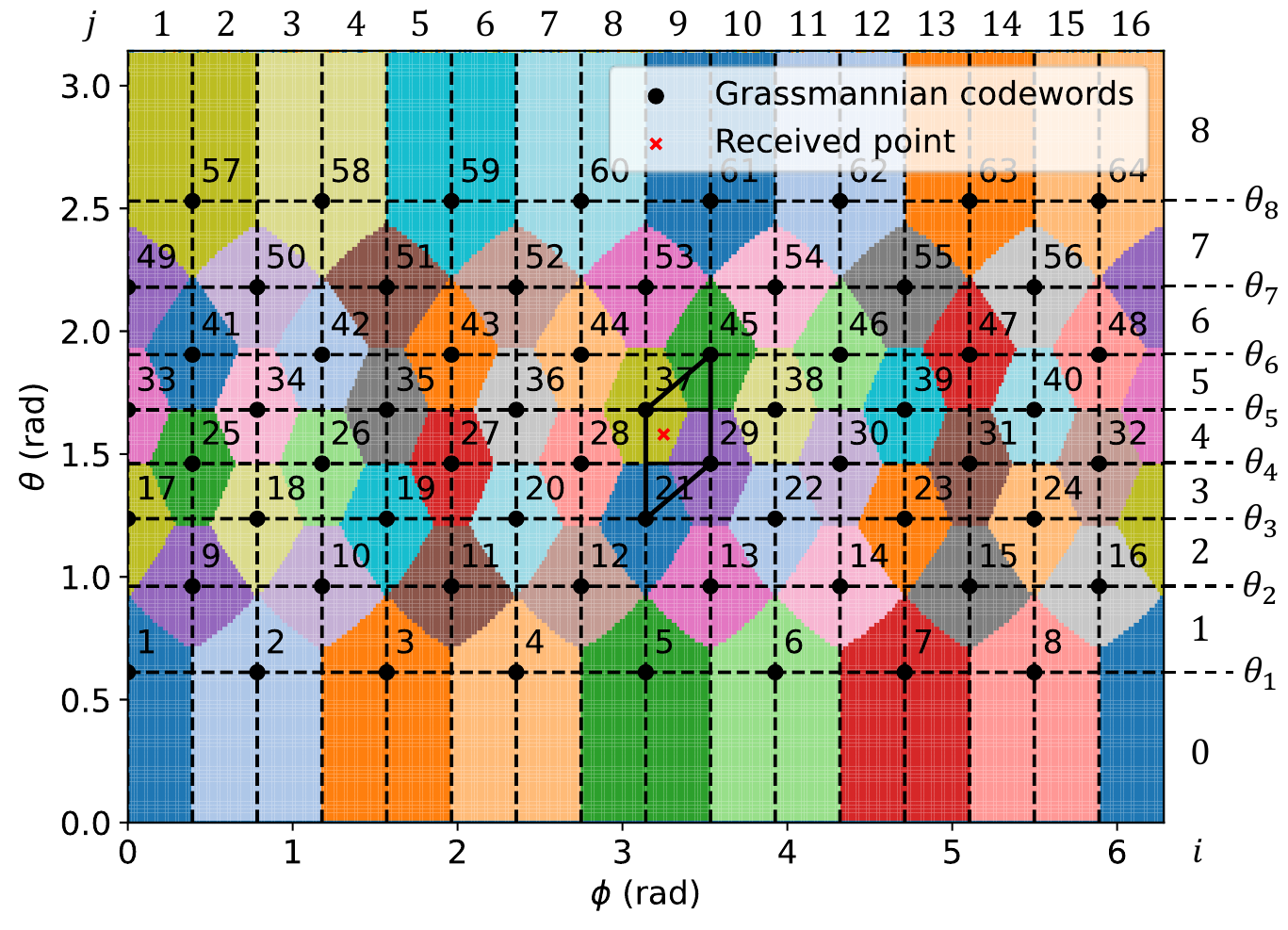}
	}
	\subfigure[
        Data representation of the nearest-neighbor regions.
        \label{subfig:index_table}
    ]{
		\includegraphics[clip, scale=0.24]{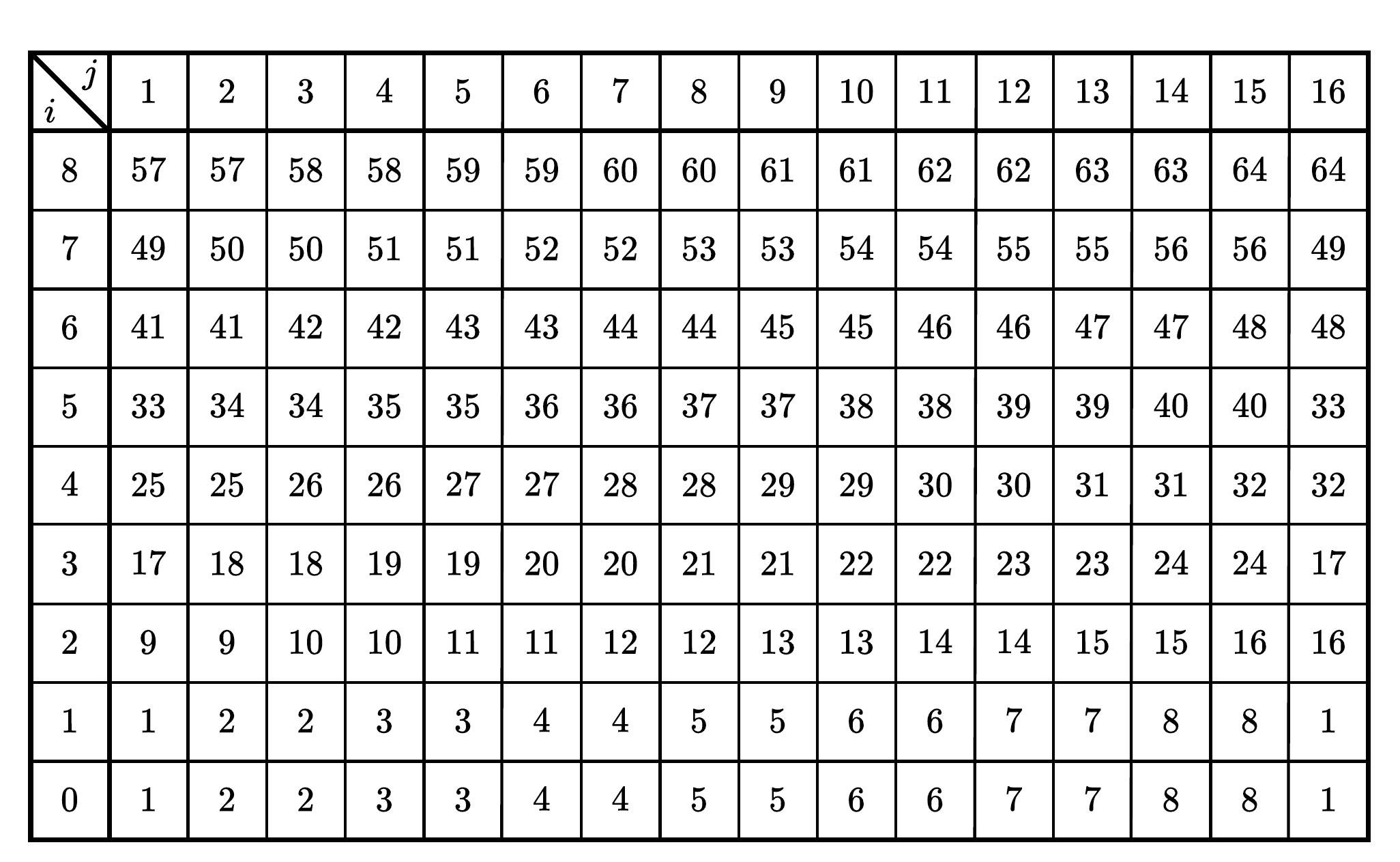}
	}
	\caption{
        A low-complexity detector using the structure diagram for $B=6$.
        \label{fig:structure_diagram=B=6}
    }
\end{figure}

The Grassmannian codewords constructed with Z-Opt can be detected efficiently using a structure diagram.
Fig. \ref{subfig:domain_B=6} shows the structure diagram for $B=6$. 
The black dots represent Grassmannian codewords, and the numbers assigned to each codeword are defined in \eqref{eq:z_opt_constellation}.
The colored regions indicate the Voronoi regions around each codeword, and the red crosses denote an example of a normalized received signal.
The vertical dashed lines correspond to \eqref{eq:phi_n}, where $j$ is the index assigned to each partitioned region.
The horizontal dashed lines correspond to \eqref{eq:Theta_n_Bneq5,7} and \eqref{eq:Theta_n_B=5,7}, where $i$ is the index assigned to each partitioned region.
By determining the grid $(i,j)$ to which the normalized received signal belongs, the set of nearest-neighbor candidates can be reduced to at most four codewords surrounding that grid within the region $j$.

In the following, the low-complexity detector is described for the case of $B\geq 4$.
For $B\in\{1,2,3\}$, the constellation size is small and the structure is simple, so these cases can be handled separately in a straightforward manner.
An overview of this low-complexity detection procedure is as follows:
\begin{enumerate}
    \item The received signal $\y$ is normalized onto the Grassmann manifold, and the polar angle $\theta_{z}$ and azimuth angle $\phi_{z}$ are computed from the normalized point.
    \item Determine a region $j$ from $\phi_{z}$.
    \item Determine a region $i$ from $\theta_{z}$ using insertion position search.
    \item From the grid $(i,j)$, identify up to four codewords that may be nearest-neighbor candidates to the received signal.
    \item Among the candidate codewords, select the one closest to the received signal.
\end{enumerate}

First, as in S-Opt, the received signal is normalized onto the Grassmann manifold using \eqref{eq:normalized_y}.
Then, using \eqref{eq:get_theta_z_and_phi_z}, the polar angle $\theta_{z}$ and azimuth angle $\phi_{z}$ of the normalized point on the Bloch sphere are obtained.

Next, the region $j$ is determined by $\phi_{z}$.
The value of $j$ can be obtained as
\begin{align}
    j=\left\lfloor \frac{\phi_{z}}{\pi/z_{\mathrm{max}}} \right\rfloor
    +1.
    \label{eq:j_hat}
\end{align}
With this computation, the nearest-neighbor candidates can already be reduced to approximately the square root order.

Subsequently, the region $i$ is determined by the following procedure.
An insertion position search is performed for $\theta_{z}$ with respect to the values $(\theta_{1},\cdots,\theta_{l})$ optimized when constructing the Z-Opt constellation.
The value of $i$ is given by
\begin{align}
    i=
    \begin{cases}
        0,&\theta_{z}<\theta_{1},\\
        1,&\theta_{1}<\theta_{z}<\theta_{2},\\
        &\vdots\\
        l-1,&\theta_{l-1}<\theta_{z}<\theta_{l},\\
        l,&\theta_{l}<\theta_{z}.
    \end{cases}
\end{align}

We then identify the nearest-neighbor candidates from the grid $(i,j)$.
First, the nearest-neighbor region, represented as in Fig.~\ref{subfig:domain_B=6}, is converted into structured data so that it can be used for detection.
For each grid $(i,j)$, we assign the index number of the codeword located on the lower edge of that grid.
Here, when $i=0$, no codeword exists on the lower edge of the grid and thus the index number of the codeword on the upper edge is assigned.
A figure in which each grid is assigned the index of its representative codeword in this manner is shown in Fig.~\ref{subfig:index_table}.
When we view this information as a function $T(i,j)$, and taking $i$ modulo $2$, for $B\notin\{5,7\}$, it is given by
\begin{align}
    T\left(i,j\right)=
    \begin{cases}
    \left( i-\delta_{j,2z_{\mathrm{max}}} \right)z_{\mathrm{max}}+\left\lfloor \frac{j}{2} \right\rfloor+1,&i=0,\\
        \left( i-1-\delta_{j,2z_{\mathrm{max}}} \right)z_{\mathrm{max}}+\left\lfloor \frac{j}{2} \right\rfloor+1,&i\equiv 1,\\
        \left( i-1 \right)z_{\mathrm{max}}+\left\lfloor \frac{j+1}{2} \right\rfloor,&\text{otherwise}.\\
    \end{cases}
\end{align}
Here, $\delta_{a,b}$ denotes the Kronecker delta, defined in general by
\begin{align}
    \delta_{a,b}=
    \begin{cases}
        1, & a=b,\\
        0, & a\neq b.
    \end{cases}
\end{align}
For $B\in\{5,7\}$, it is given by
\begin{align}
    T\left(i,j\right)=\left( i-\frac{3}{2} \right)z_{\mathrm{max}}+t\left( i,j \right),
\end{align}
where
\begin{align}
    &t\left( i,j \right)=
    &\begin{cases}
        \lfloor \frac{j+1}{4} \rfloor+1+\frac{1}{2}z_{\mathrm{max}}\Delta\left(i,j\right),&i\in \left\{0,1,l\right\},\\
        \lfloor \frac{j+1}{2} \rfloor,&i\neq 0\;\text{and}\;i\equiv 0,\\
        \lfloor \frac{j}{2} \rfloor +1-z_{\mathrm{max}}\delta_{j,2z_{\mathrm{max}}},&\text{otherwise},
    \end{cases}
\end{align}
and
\begin{align}
    \Delta\left( i,j \right)=3\delta_{i,0}+\delta_{i,1}-\delta_{j,2z_{\mathrm{max}}}-\delta_{j,2z_{\mathrm{max}}-1}.
\end{align}

Next, within region $j$, the region $\hat{i_{c}}$ that minimizes the chordal distance to the received signal is obtained as
\begin{align}
    \hat{i_{c}}=\operatorname*{argmin}_{
    {
    i_{c}\in\mathcal{I}_{c}
    }
    }
    {
    d_{\mathrm{d}}\left(\theta_{i_{c}},\theta_{z},
    \Delta\phi\left(i_{c},j\right)
    \right)
    },
    \label{eq:i_hat}
\end{align}

where $\mathcal{I}_{c}=\left\{\;i_{c} \;\middle|\; \max{(0,i-1)}\leq i_{c}\leq\min{(i+2,l)}  \;\right\}$.
When $B\notin \{5,7\}$, $\Delta\phi(i_{c},j)$ is given by
\begin{align}
    \Delta\phi\left(i_{c},j\right)=\left|
    \phi_{z}-(j-f(i_{c},j))\frac{\pi}{z_{\mathrm{max}}}
    \right|,
\end{align}
where $f(i,j)$ is defined modulo 2 as
\begin{align}
    f\left(i,j\right)=
    \begin{cases}
        1, &(i+\delta_{i,0})\equiv j,\\
        0, &\text{otherwise}.
    \end{cases}
\end{align}
When $B\in \{5,7\}$, $\Delta\phi(i_{c},j)$ is given by
\begin{align}
    \Delta\phi\left(i_{c},j\right)=
    \begin{cases}
        \left|
        \phi_{z}-(j-g(j))\frac{\pi}{z_{\mathrm{max}}}
        \right|,
        & i_{c}\in\left\{0,1,l\right\},
        \\[0.5em]
        \left|
        \phi_{z}-(j-f(i_{c},j))\frac{\pi}{z_{\mathrm{max}}}
        \right|,
        & \text{otherwise},
    \end{cases}
\end{align}
where $g(j)$ is defined modulo 4 as
\begin{align}
    g\left(j\right)=
    \begin{cases}
         1, & j\equiv 1, \\
         2, & j\equiv 2, \\
        -1, & j\equiv 3, \\
         0, & \text{otherwise}. \\
    \end{cases}
    \label{eq:g_function}
\end{align}
Finally, by computing $T(\hat{i_{c}},j)$ from the values of $\hat{i_{c}}$ and $j$ obtained from \eqref{eq:i_hat} and \eqref{eq:j_hat}, respectively, the index number of the transmitted symbol can be determined.

The exactness of the Z-Opt detector is summarized in the following proposition.

\begin{proposition}[Exactness of the Z-Opt Detector]
    For the Z-Opt constellation with $B\ge 4$ and any received point on the Bloch sphere, the candidate set generated by \eqref{eq:j_hat}--\eqref{eq:g_function} contains a nearest Grassmannian codeword.
    Moreover, this candidate set contains at most four codewords.
    Therefore, selecting the closest codeword from this set produces the same decision as exhaustive GLRT detection.
    \label{prop:exactness}
\end{proposition}

The proof is provided in Appendix~\ref{appendix:proof_of_z-opt_detector}.
It first restricts the search to the relevant azimuthal region and then shows that at most four points within that region can be nearest-neighbor candidates.
The special structures used for \(B\in\{5,7\}\) are handled separately through virtual points.

The GLRT detector performs exhaustive search--based detection by computing the distances to all $C$ codewords and selecting the one with the maximum likelihood.
Therefore, the receiver must store all $C$ codewords.
In contrast, the Z-Opt detector first narrows down the candidates to at most four codewords using $O(\log_{2}{\sqrt{C}})$ numerical comparisons and then computes the distances only for these candidates.
Moreover, detection can be performed without storing all codewords.
It suffices to store $O(\sqrt{C})$ values, denoted by $\theta_{1}, \ldots, \theta_{l}$,
at the receiver.

\section{Performance Results}
\label{sec:performance_results}
In this section, we compare the proposed methods with conventional methods.
Here, the constellation proposed in Section \ref{sec:s-opt_constellation} is referred to as the S-Opt constellation, and its low-complexity detector as the S-Opt detector.
The constellation proposed in Section \ref{sec:z-opt_constellation} is referred to as the Z-Opt constellation, and its low-complexity detector as the Z-Opt detector.
We compare the constellations in terms of minimum chordal distance and symbol error rate (SER).
The low-complexity detectors are compared in terms of time complexity, space complexity, and SER.
In addition, regarding the construction cost of the constellations, we compare the number of distance computations in the objective function and the number of optimization variables.

We consider the system model given in \eqref{eq:system_model}, where the elements of $\H$ and $\W$ follow independent complex Gaussian distributions of $\mathcal{CN}(0,1)$ and $\mathcal{CN}(0,\sigma^{2})$, respectively.

\subsection{Constellation Performance}
\begin{figure}[tb]
	\centering
    \includegraphics[clip, scale=0.68]{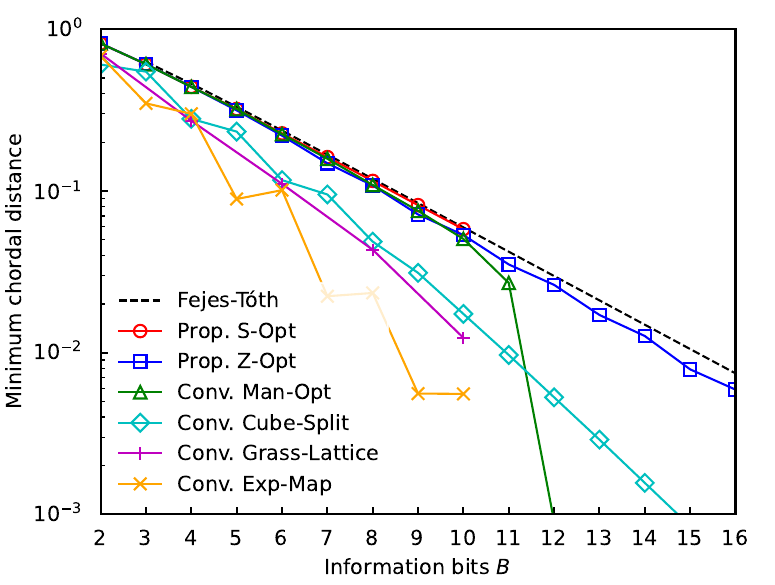}
	\caption{
        Comparison of the minimum chordal distance for each constellation.
    \label{fig:minimum_chordal_distance}
    }
\end{figure}

Fig.~\ref{fig:minimum_chordal_distance} compares the minimum chordal distance for each constellation.
The Fejes--T\'oth bound is used as the reference upper bound on the minimum chordal distance, as given in \eqref{eq:fejes_toth}.
It can be observed that S-Opt achieves the theoretical limit corresponding to the evaluated optimal configurations.
Although Z-Opt does not outperform S-Opt, it approaches the same bound.
Since the objective function in Man-Opt becomes exponentially more complex as $B$ increases, it becomes difficult to search for good solutions, and the performance degrades for $B\geq 10$.

\begin{figure}[tb]
	\centering
    \includegraphics[clip, scale=0.68]{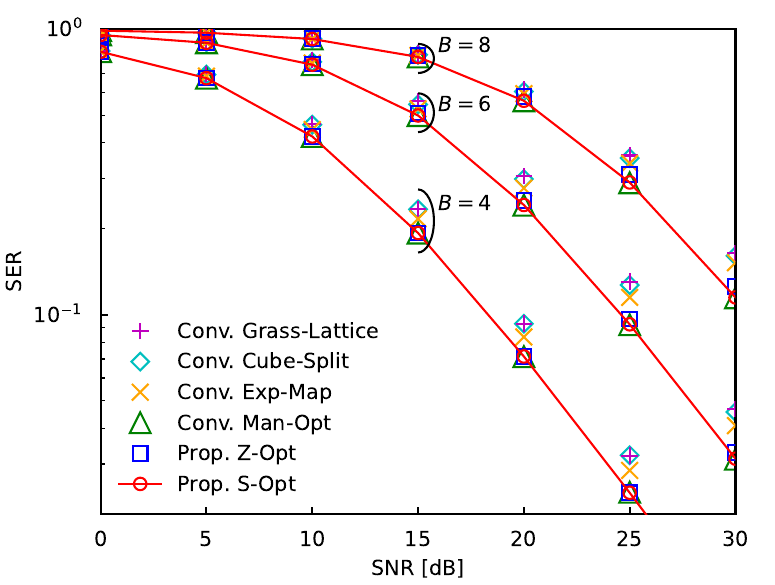}
	\caption{
        Comparison of SERs for each constellation at $B\in \{4,6,8\}$.
    \label{fig:SNR_vs_SER}
    }
\end{figure}

Fig.~\ref{fig:SNR_vs_SER} compares the SERs for each constellation at $B\in\{4,6,8\}$.
The minimum chordal distance is strongly correlated with the SER, and a larger minimum chordal distance generally results in a lower SER.
Therefore, S-Opt and Man-Opt, which have the largest minimum chordal distances, exhibit the lowest SERs, followed by Z-Opt.

\subsection{Reduction of Complexity and Optimality}
Table~\ref{table:comparison_complexity} compares the time complexity and space complexity of GLRT, S-Opt, and Z-Opt detectors.
The time complexity is evaluated in terms of the number of complex multiplications, while the space complexity is evaluated by the number of floating-point values stored at the receiver.

In the GLRT detector, each matrix multiplication of $\Y^{\herm}\X$ requires $NMT$ complex multiplications.
This matrix multiplication is performed $C$ times per detection, and the total number of complex multiplications is $NMTC$.
Since $M=1$ and $T=2$ are fixed, the time complexity is $O(NC)$.
Regarding space complexity, the receiver stores the entire constellation, resulting in a space complexity of $O(C)$.

In the S-Opt detector, there are two main processes that dominate the order of the time complexity.
The first process is the SVD performed during the rough estimation when $N \geq 2$.
The SVD of $\Y \in \mathbb{C}^{T \times N}$ requires a computational complexity of $O(T^{2}N + T^{3})$ when $N \geq T$. Since $T=2$ is fixed, this reduces to $O(N)$.
The second process is the nearest-neighbor search using a KD-tree. The computational complexity of a KD-tree search is $O(D \log_{2} C)$, where $D$ denotes the dimension of the Euclidean space.
Because the Euclidean dimension of the Bloch sphere is fixed at $D = 3$, the complexity becomes $O(\log_{2} C)$.
Combining the rough estimation and the KD-tree nearest-neighbor search, the overall time complexity of the S-Opt detector is $O(N+\log_{2}{C})$.
In terms of space complexity, the constructed KD-tree requires $O(C)$, and since the receiver must also store the entire constellation, the overall space complexity remains $O(C)$.

In the Z-Opt detector, the only process that significantly affects the order of the time complexity is the SVD.
Therefore, similar to the S-Opt detector, the time complexity is $O(N)$.
As for space complexity, detection can be performed by storing only the values of $\theta$, resulting in a space complexity of $O(\sqrt{C})$.

In addition, the number of distance metric evaluations and numerical comparisons per detection is also examined for each detector.
In the GLRT detector, the distance metric $\lVert \Y^{\herm}\X \rVert_{\mathrm{F}}$ is computed $C$ times, and $C$ numerical comparisons are performed to find the maximum value.
In the S-Opt detector, since nearest-neighbor detection is carried out using a KD-tree, both the number of distance metric evaluations and the number of numerical comparisons are $O(\log_{2} C)$.
For the Z-Opt detector, candidate transmit signals are first narrowed down, and at most four chordal distance computations are required.
Therefore, the number of distance metric evaluations is $O(1)$.
Regarding numerical comparisons, determining region $i$ involves an insertion position search over an ordered set of $O(\sqrt{C})$ values of $\theta$, resulting in a complexity of $O(\log_{2} \sqrt{C})$.

\begin{table}[tb]
    \begin{center}
    \renewcommand{\arraystretch}{1.0}
    \setlength{\tabcolsep}{5pt}
    \caption{Comparison of detector computational complexity}
    \label{table:comparison_complexity}
    \begin{tabular}{|c|c|c|}
        \hline
        Detector & Time complexity    & Space complexity \\ \hline
        GLRT     & $O(NC)$            & $O(C)$           \\ \hline
        S-Opt    & $O(N+\log_{2}{C})$ & $O(C)$           \\ \hline
        Z-Opt    & $O(N)$             & $O(\sqrt{C})$    \\
        \hline
    \end{tabular}
    \end{center}
\end{table}

\begin{figure}[tb]
	\centering
    \includegraphics[clip, scale=0.68]{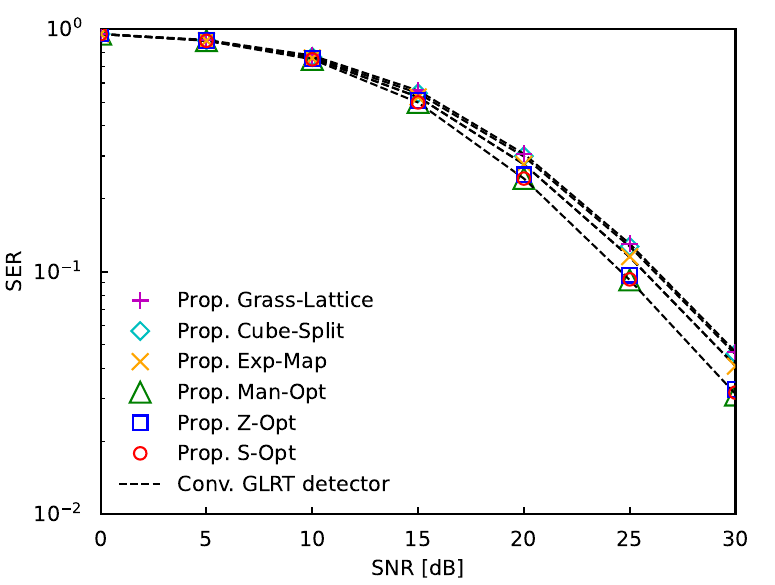}
	\caption{
        SER comparison between the GLRT and S-Opt detectors at $B=6$.
        Markers indicate S-Opt detection, while dashed lines correspond to GLRT detection. \label{fig:GLRT_vs_KD-Tree}
    }
\end{figure}

Fig.~\ref{fig:GLRT_vs_KD-Tree} compares the SERs between the GLRT and S-Opt detectors for each constellation at $B=6$.
It can be observed that the S-Opt detector achieves the same detection accuracy as the GLRT detector for arbitrary constellations.

\begin{figure}[tb]
	\centering
    \includegraphics[clip, scale=0.68]{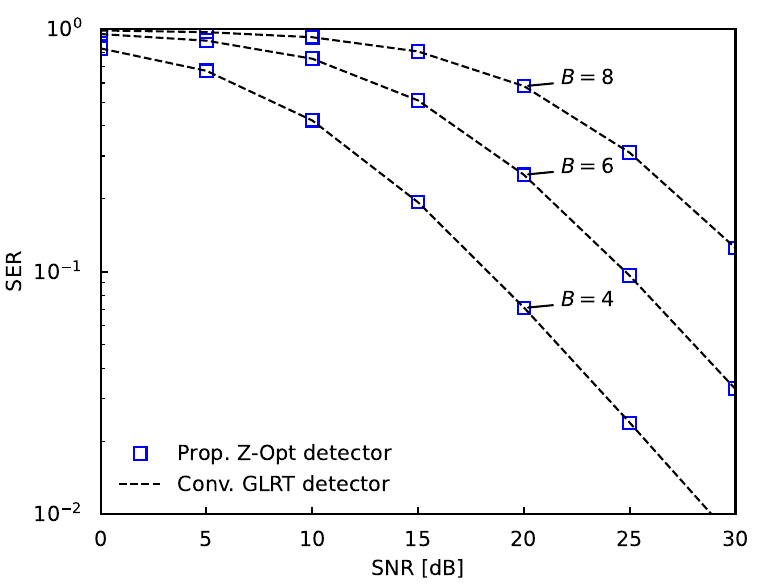}
	\caption{
        SER comparison between the GLRT and Z-Opt detectors for the Z-Opt constellation at $B\in\{4,6,8\}$.
    \label{fig:GLRT_vs_Z-Opt}
    }
\end{figure}

Fig.~\ref{fig:GLRT_vs_Z-Opt} compares the SERs of GLRT and Z-Opt detectors for the Z-Opt constellation at $B=4,6$ and $8$.
For the Z-Opt constellation, the Z-Opt detector achieves the same detection accuracy as the GLRT detector.
A proof that the Z-Opt detector always yields the same detection result as the GLRT detector is provided in Appendix~\ref{appendix:proof_of_z-opt_detector}.

\subsection{Construction Cost}
As described in Section~\ref{sec:s-opt_constellation}, since the S-Opt constellation is constructed using existing optimal configurations and optimized results, the construction cost can be regarded as negligible.

\begin{figure}[tb]
	\centering
    \includegraphics[clip, scale=0.68]{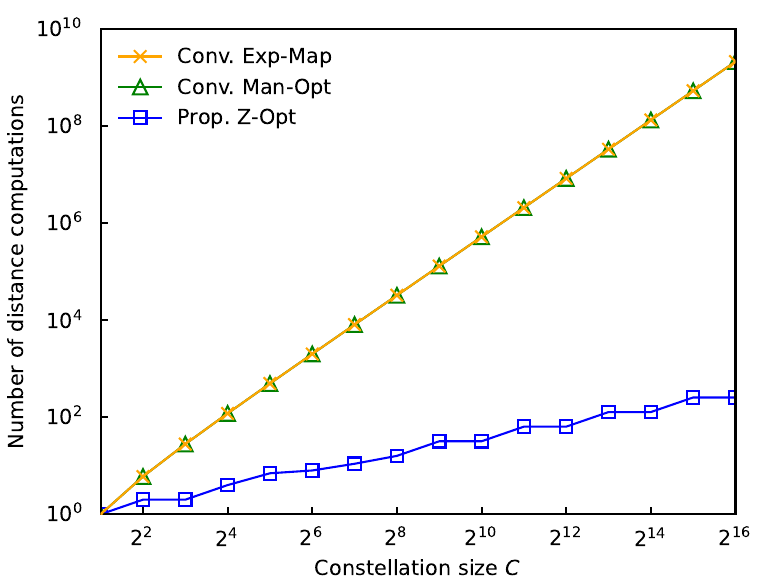}
	\caption{
        Distance-metric computations per objective-function evaluation for Exp-Map, Man-Opt, and Z-Opt.
    \label{fig:comparison_evaluation}
    }
\end{figure}

Fig.~\ref{fig:comparison_evaluation} compares the number of distance-metric computations per objective-function evaluation among Man-Opt, Exp-Map, and Z-Opt.
Man-Opt is defined by \eqref{eq:Man-Opt_logsumexp} and computes $\binom{C}{2}=C(C-1)/2$ distance metrics per objective function evaluation.
Therefore, its computational order is $O(C^{2})$.
In Exp-Map, the minimum chordal distance can be improved by adjusting the radii of the symbols before constructing the constellation.
Since this adjustment is performed by maximizing the minimum chordal distance of the constructed constellation as the objective function, the required number of distance metric computations is the same as that of Man-Opt.
In contrast, Z-Opt requires the computation of $2n_{v}$ distance metrics when $B\notin \{5,7\}$, and $2n_{v}+3$ distance metrics when $B\in \{5,7\}$.
Since $n_{v}$ is $O(\sqrt{C})$, the number of distance metric computations required by Z-Opt is of order $O(\sqrt{C})$.

\begin{figure}[tb]
	\centering
    \includegraphics[clip, scale=0.68]{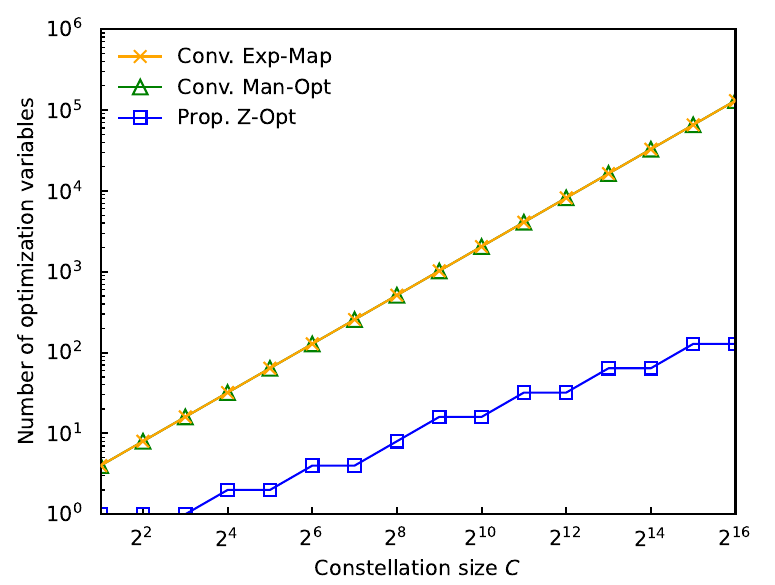}
	\caption{
        The number of optimization variables for Exp-Map, Man-Opt, and Z-Opt.
    \label{fig:number_of_variables}
    }
\end{figure}

Fig.~\ref{fig:number_of_variables} compares the number of optimization variables among Exp-Map, Man-Opt, and Z-Opt.
In Man-Opt, a codeword on the Grassmann manifold is parameterized by $2M(T-M)$ real numbers and optimized, so the number of optimization variables is $2C$.
For Exp-Map, when $M=1$ and $T=2$, complex numbers whose magnitudes before the exponential map are within $\pi/2$ correspond one-to-one with points on the Grassmann manifold.
Hence, optimization on the Grassmann manifold can be performed by optimizing the symbols before the exponential map.
Because each complex number consists of two real numbers, the number of optimization variables is also $2C$.
In contrast, Z-Opt only requires the optimization of $n_{v}$ parameters $\theta$, and its order is $O(\sqrt{C})$.

\section{Conclusion}
\label{sec:conclusion}
In this paper, we presented two construction methods for Grassmannian constellations of one-dimensional subspaces in a two-dimensional space on $\mathcal{G}(2,1)$, termed S-Opt and Z-Opt, and proposed two low-complexity detectors.
Both the construction and detection procedures are performed on the Bloch sphere.
The S-Opt constellation achieves the theoretical limit corresponding to the considered optimal Bloch-sphere packings.
The S-Opt detector can be applied to general Grassmannian constellations on $\mathcal{G}(2,1)$ and achieves the same detection performance as the GLRT detector.
It requires $O(N+\log_{2}{C})$ complex multiplications, $O(\log_{2}{C})$ numerical comparisons, and $O(C)$ space complexity.
The Z-Opt constellation achieves a minimum chordal distance that approaches the derived upper bound with low-complexity optimization.
The Z-Opt detector is applicable only to the specific Z-Opt constellation and achieves the same detection performance as the GLRT detector.
It requires $O(N)$ complex multiplications, $O(\log_{2}{\sqrt{C}})$ numerical comparisons, and $O(\sqrt{C})$ space complexity.
In addition, we derived a theoretical upper bound on the minimum chordal distance based on the Fejes--T\'oth bound.
Future work includes extending these construction methods and low-complexity detectors to general values of $M$ and $T$.

\appendices
\section{Exactness of the Z-Opt Detector}
\label{appendix:proof_of_z-opt_detector}
We prove that the Z-Opt detector always yields the same detection result as the GLRT detector when the Z-Opt constellation is employed.
Specifically, we show that, for any received signal, the candidate set generated by the Z-Opt detector always contains the true nearest Grassmannian codeword while consisting of at most four candidates.
We first introduce several properties of distances on the unit sphere that are used throughout the proof.
We then present the proof for the case of $B\notin\{5,7\}$.
Finally, we present the proof for the case of $B\in\{5,7\}$ and introduce the concept of virtual points, which are used to simplify the detection algorithm in this case.
As in Section~\ref{sec:z-opt_constellation}, only the case of $B\geq 4$ is considered.

\subsection{Preliminaries}
We first introduce properties of distances on the unit sphere that are used throughout the proof.
Let the spherical coordinates of a point $s$ on the unit sphere be denoted by $(\theta_{s},\phi_{s})$.
Let $d_{\mathrm{E}}(s_{i},s_{j})$ denote the Euclidean distance between two points $s_{i}$ and $s_{j}$.
Since the azimuthal angle $\phi$ has a period of $2\pi$, we define the angular difference between two points as
\begin{align}
    \Delta\phi\left(s_{i},s_{j}\right)=
    \min{
    \left\{
    \left|
    \phi_{s_{i}}-\phi_{s_{j}}
    \right|,
    2\pi-
    \left|
    \phi_{s_{i}}-\phi_{s_{j}}
    \right|
    \right\}
    }.
\end{align}
The following two lemmas hold.
\begin{lemma}
    [Nearest Point on a Common Latitude]
    Suppose that three points $a$, $b$, and $c$ lie on the unit sphere, with $\theta_{a}=\theta_{b}$.
    The ordering of $d_{\mathrm{E}}(a,c)$ and $d_{\mathrm{E}}(b,c)$ coincides with that of $\Delta\phi(a,c)$ and $\Delta\phi(b,c)$.
    \label{lem:same_theta}
\end{lemma}
\begin{proof}
    The result follows immediately from \eqref{eq:dd}.
\end{proof}
\begin{lemma}
    [Boundary between Vertically Aligned Points]
    Suppose that two points $a$ and $b$ lie on the unit sphere, with a common azimuth angle $\phi_{a}=\phi_{b}=\phi_{0}$.
    Let $\Delta\phi=\phi-\phi_0$.
    Then, when $|\phi-\phi_{0}|\ll 1$, the boundary of the spherical Voronoi regions of $a$ and $b$ can be approximated by
    \begin{align}
        \theta\simeq
        \frac{\theta_{a}+\theta_{b}}{2}.
    \end{align}
    \label{lem:same_phi}
\end{lemma}
\begin{proof}
    The coordinates of points $a$ and $b$ are given by
    \begin{align}
        \mathbf{a}=
        \begin{pmatrix}
            \sin{\theta_{a}}\cos{\phi_{0}} \\
            \sin{\theta_{a}}\sin{\phi_{0}} \\
            \cos{\theta_{a}}
        \end{pmatrix}
        ,
        \mathbf{b}=
        \begin{pmatrix}
            \sin{\theta_{b}}\cos{\phi_{0}} \\
            \sin{\theta_{b}}\sin{\phi_{0}} \\
            \cos{\theta_{b}}
        \end{pmatrix}.
    \end{align}
    Let the coordinates of an arbitrary point $s$ on the unit sphere be
    \begin{align}
        \mathbf{s}=
        \begin{pmatrix}
            \sin{\theta}\cos{\phi} \\
            \sin{\theta}\sin{\phi} \\
            \cos{\theta}
        \end{pmatrix}.
    \end{align}
    Then, the condition that $s$ is equidistance from $a$ and $b$ is
    \begin{align}
        \mathbf{s}\cdot\left(\mathbf{a}-\mathbf{b}\right)=0.
    \end{align}
    Solving this equation for $\theta$ yields
    \begin{align}
        \theta=
        \tan^{-1}{
        \left(
        \frac{
        \tan{
        \left(
        \frac{\theta_{a}+\theta_{b}}{2}
        \right)
        }
        }{
        \cos{
        \left(
        \phi-\phi_{0}
        \right)
        }
        }
        \right)
        }.
    \end{align}
    Let $\Delta\phi=\phi-\phi_{0},\;\alpha=(\theta_{a}+\theta_{b})/2$.
    Then,
    \begin{align}
        \theta=
        \tan^{-1}{
        \left(
        \frac{
        \tan{\alpha}
        }{
        \cos{\Delta\phi}
        }
        \right)
        }
        \label{eq:theta_function}
    \end{align}
    holds.
    When $|\Delta\phi|\ll 1$, the Taylor expansion of $\cos\Delta\phi$ about $\Delta\phi=0$ is
    \begin{align}
        \cos{\Delta\phi}=
        1-\frac{\Delta\phi^{2}}{2}+
        O\left(\Delta\phi^{4}\right).
    \end{align}
    Similarly, the Taylor expansion of $1/(1-t)$ about $t=0$ is
    \begin{align}
        \frac{1}{1-t}=
        1+t+O\left(t^{2}\right).
    \end{align}
    Substituting $t=\Delta\phi^{2}/2+O(\Delta\phi^{4})$ gives
    \begin{align}
        \frac{1}{\cos{\Delta\phi}}=
        1+\frac{
        \Delta\phi^{2}
        }{
        2
        }
        +O\left(
        \Delta\phi^{4}
        \right).
    \end{align}
    Hence,
    \begin{align}
        \frac{\tan{\alpha}}{\cos{\Delta\phi}}=
        \tan{\alpha}+\frac{\tan{\alpha}}{2}\Delta\phi^{2}
        +O\left(
        \Delta\phi^{4}
        \right)
    \end{align}
    Substituting this into \eqref{eq:theta_function} yields
    \begin{align}
        \theta=
        \tan^{-1}{
        \left(
        \tan{\alpha}+\epsilon
        \right)
        },
    \end{align}
    where
    \begin{align}
        \epsilon=
        \frac{\tan{\alpha}}{2}\Delta\phi^{2}+
        O\left(
        \Delta\phi^{4}
        \right).
    \end{align}
    Since $\epsilon=O(\Delta\phi^{2})$, the Taylor expansion of $\tan^{-1}{(t+\epsilon)}$ about $t=\tan{\alpha}$ gives
    \begin{align}
        \tan^{-1}{
        \left(
        \tan{\alpha}+\epsilon
        \right)
        }=
        \alpha+\frac{\epsilon}{1+\tan^{2}{\alpha}}
        +O\left(\epsilon^{2}\right)
    \end{align}
    Substituting the expression for $\epsilon$ and simplifying, we obtain
    \begin{align}
        \theta=
        \alpha+
        \frac{\sin{2\alpha}}{4}
        \Delta\phi^{2}
        +O\left(
        \Delta\phi^{4}
        \right).
    \end{align}
    Therefore, when $|\Delta\phi|\ll1$, the correction term proportional to $\Delta\phi^2$ is negligible, and
    \begin{align}
        \theta\simeq
        \alpha=\frac{
        \theta_{a}+\theta_{b}
        }{2}
    \end{align}
    is obtained.
\end{proof}

\subsection{Case of $B\notin\{5,7\}$}
We next prove that, when $B\notin\{5,7\}$, there are at most four nearest-neighbor candidates for any received signal.

\begin{figure}[tb]
    \centering
    \includegraphics[clip, scale=0.50]{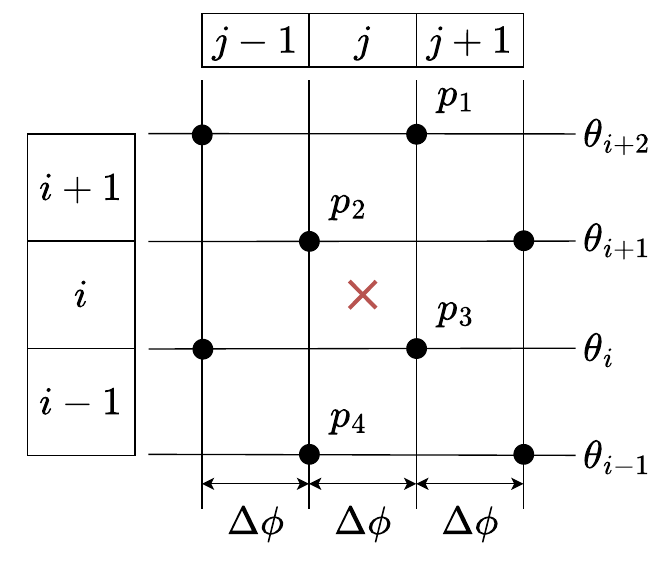}
    \caption{
        Example of the received signal and its nearest-neighbor candidates for $B\notin\{5,7\}$.
        \label{fig:Bnotin57_appendix}
    }
\end{figure}

We first show that, when the received signal lies in region $(i,j)$, the nearest-neighbor candidates are restricted to the points in region $j$.
As illustrated in Fig.~\ref{fig:Bnotin57_appendix}, the Z-Opt constellation has a regular structure in which, for every value of $\theta$, there always exists a point on either the left or right boundary of region $j$.
Moreover, for each value of $\theta$, the points are equally spaced with an interval of $2\Delta\phi=2\pi/z_{\mathrm{max}}$.
By Lemma~\ref{lem:same_theta}, among the points having the same $\theta$, the one with the smallest distance to the received signal is the point whose $\phi$ is the closest to that of the received signal.
Such a point always belongs to region $j$.

Therefore, when the received signal lies in region $(i,j)$, only the points in region $j$ need to be considered as nearest-neighbor candidates.

Next, we show that there are at most four nearest-neighbor candidates in region $j$.
As shown in Fig.~\ref{fig:Bnotin57_appendix}, the boundary between the spherical Voronoi regions of $p_{1}$ and the point located above $p_{1}$ with the same $\phi$ lies between the two points within region $j$, according to Lemma~\ref{lem:same_phi}.
Hence, the received signal always belongs to the Voronoi region of $p_{1}$.
The same argument applies to $p_{2}$, $p_{3}$, and $p_{4}$.
Therefore, when the received signal lies in region $(i,j)$, there are at most four nearest-neighbor candidates in region $j$.

When the received signal belongs to region $i=0$, there are no points corresponding to $p_{3}$ and $p_{4}$, and thus only two candidates remain.
Similarly, when $i=1$, there is no point corresponding to $p_{4}$, leaving three candidates.
By symmetry, there are three candidates when $i=l-1$ and two candidates when $i=l$.

Consequently, when $B\notin\{5,7\}$ and the received signal lies in region $(i,j)$, there are two nearest-neighbor candidates in region $j$ for $i=0$ or $l$, three candidates for $i=1$ or $l-1$, and four candidates otherwise.
Therefore, the number of nearest-neighbor candidates is at most four.

\subsection{Case of $B\in\{5,7\}$}
\begin{figure}[tb]
    \centering
    \includegraphics[clip, scale=0.50]{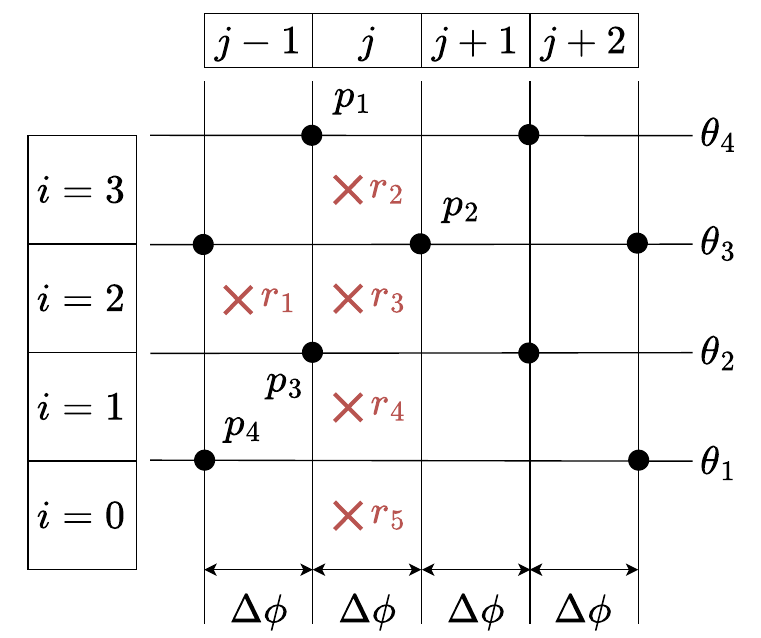}
    \caption{
        Example of the received signal and its nearest-neighbor candidates for $B\in\{5,7\}$.
        \label{fig:Bin57_appendix_a}
    }
\end{figure}

When $B\in\{5,7\}$, as illustrated in Fig.~\ref{fig:Bin57_appendix_a}, the number of points on $\theta_{1}$ and $\theta_{l}$ is half that on the other values of $\theta$, and the spacing between adjacent points is $4\Delta\phi$.
As a result, there are cases in which the nearest-neighbor candidates differ from those in the case of $B\notin\{5,7\}$.

For example, when the received signal is located at $r_{1}$ or $r_{2}$ in Fig.~\ref{fig:Bin57_appendix_a}, the nearest-neighbor candidates are, as in the case of $B\notin\{5,7\}$, at most four points in region $j$.
However, when the received signal is located at $r_{3}$, $r_{4}$, or $r_{5}$, there is no point on $\theta_{1}$ in region $j$.
We therefore show that, even in these cases, there are at most four nearest-neighbor candidates.

First, in Fig.~\ref{fig:Bin57_appendix_a}, we ignore the points on $\theta_{1}$ and determine the nearest-neighbor candidates.
The remaining points have the same structure as in the case of $B\notin\{5,7\}$.
Therefore, the nearest-neighbor candidates are restricted to the points in region $j$.
For example, the candidates are $p_{1}$, $p_{2}$, and $p_{3}$ for $r_{3}$, $p_{2}$ and $p_{3}$ for $r_{4}$, and $p_{2}$ and $p_{3}$ for $r_{5}$.
We then add the point on $\theta_{1}$ whose $\phi$ is closest to that of the received signal.
That is, by adding $p_{4}$ to the candidates in region $j$, the numbers of candidates become four for $r_{3}$, three for $r_{4}$, and three for $r_{5}$.

For $r_{5}$, only $p_{3}$ and $p_{4}$ are regarded as nearest-neighbor candidates, excluding $p_{2}$.
When $\Delta\phi\ll 1$, \eqref{eq:dd} can be approximated by \eqref{eq:dv}, and the distances from the received signal to $p_{2}$ and $p_{3}$ are determined solely by the difference in $\theta$.
Since $p_{3}$ is closer to $r_{5}$ than $p_{2}$, only $p_{3}$ and $p_{4}$ are considered as nearest-neighbor candidates.

Consequently, when $B\in\{5,7\}$ and the received signal lies in region $(i,j)$, there are two nearest-neighbor candidates for $i=0$ or $l$, three candidates for $i=1$ or $l-1$, and four candidates otherwise.
Therefore, the number of nearest-neighbor candidates is at most four.
It also follows that the number of candidates for each region $i$ is identical to that in the case of $B\notin\{5,7\}$.
Since the candidate set always contains the true nearest Grassmannian codeword for $B\geq 4$, selecting the nearest Grassmannian codeword from this candidate set yields the same detection result as the GLRT detector.

\begin{figure}[tb]
    \centering
    \includegraphics[clip, scale=0.50]{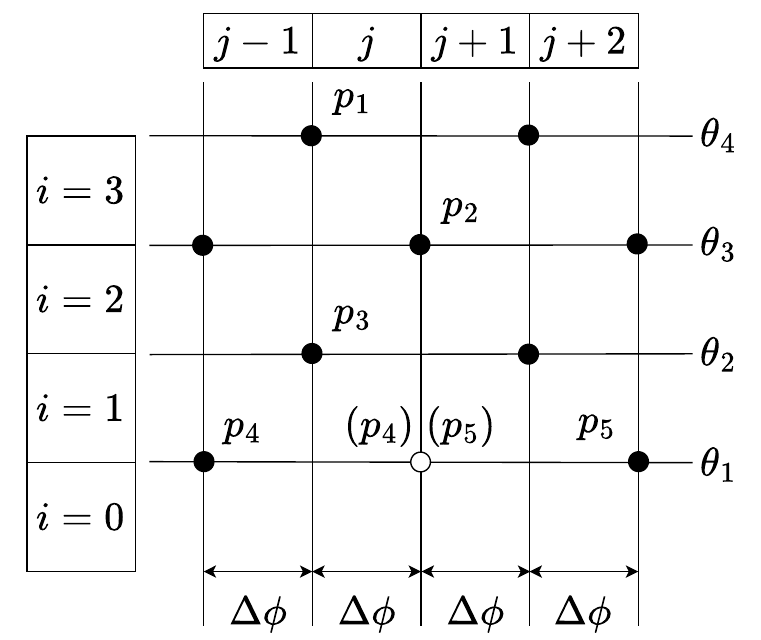}
    \caption{
    Example of virtual points for $B\in\{5,7\}$.
    \label{fig:Bin57_appendix_b}
    }
\end{figure}

Furthermore, when $B\in\{5,7\}$, the same structure as that for $B\notin\{5,7\}$ can be reproduced by introducing virtual points.
The virtual point shown by an open circle in Fig.~\ref{fig:Bin57_appendix_b} is regarded as $p_{4}$ in region $j$ and as $p_{5}$ in region $j+1$.
In the Z-Opt detector, the nearest-neighbor candidates are determined in the same manner as for $B\notin\{5,7\}$ by assuming the existence of the virtual points.
As a result, the obtained candidates coincide with those derived above for the case of $B\in\{5,7\}$.

\footnotesize{
	\bibliographystyle{IEEEtranURLandMonthDeactivated}
	\bibliography{main}
}

\begin{IEEEbiography}[{\includegraphics[width=1in,height=1.25in,clip,keepaspectratio]{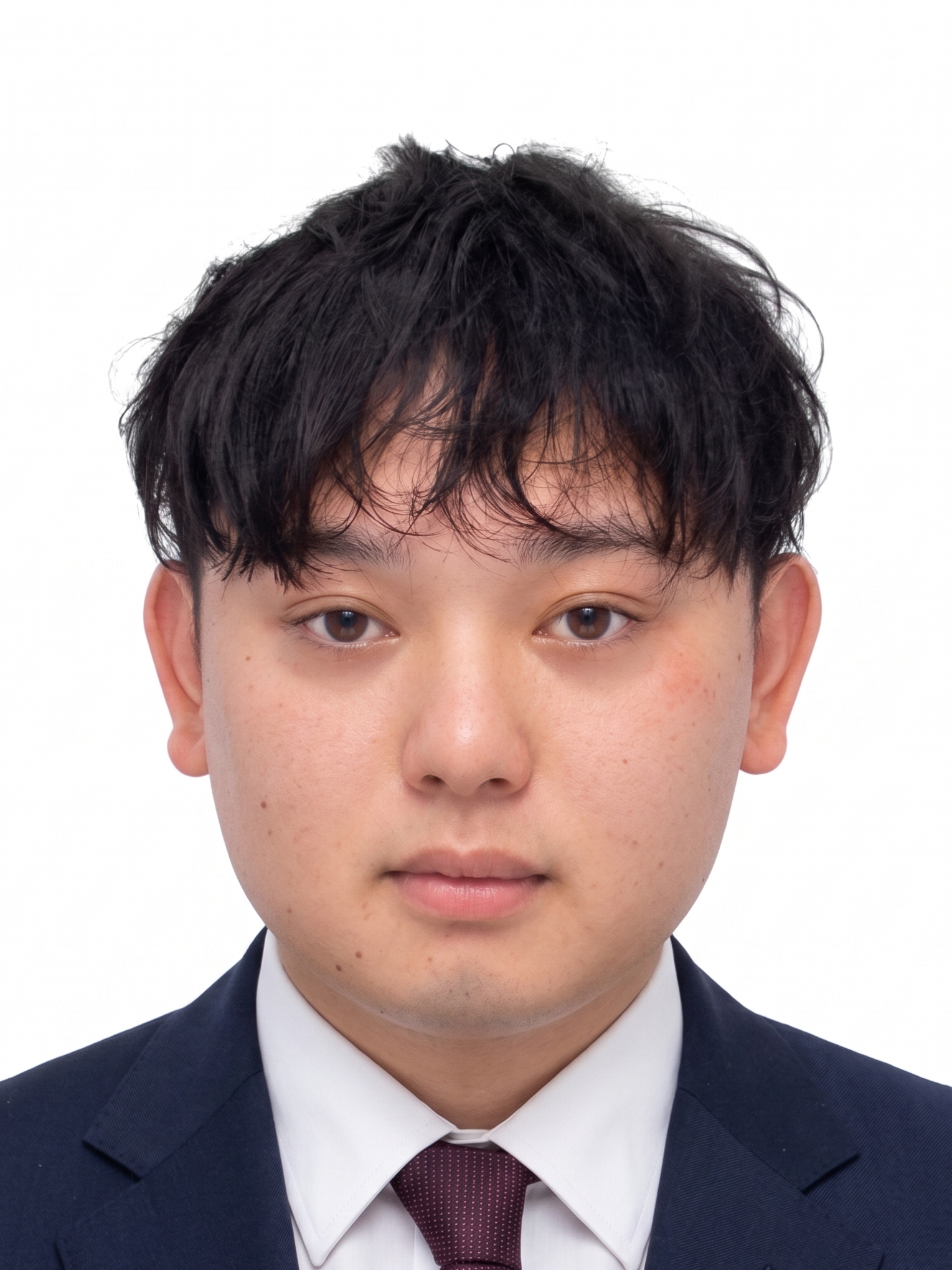}}]{Kotaro Shigenaga}~(Graduate Student Member, IEEE) received the B.E. degree from Yokohama National University, Yokohama, Japan, in 2025, where he is currently pursuing the M.E. degree with the Department of Mathematics, Physics, Electrical Engineering and Computer Science, Graduate School of Engineering Science. His research interests include noncoherent communications.
\end{IEEEbiography}

\begin{IEEEbiography}[{\includegraphics[width=1in,height=1.25in,clip,keepaspectratio]{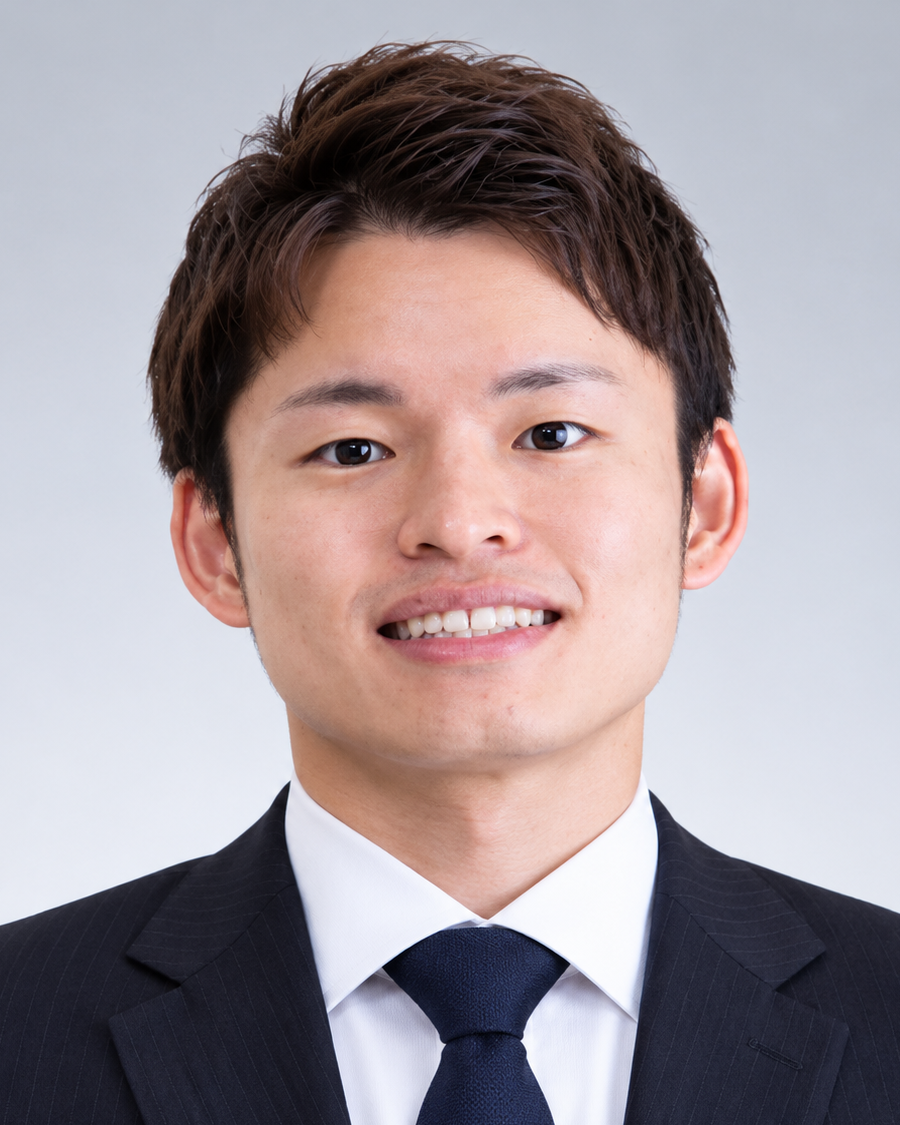}}]{Hiroki Iimori} 
~(Member, IEEE) is a Senior Researcher at Ericsson Research, working on 3GPP MIMO technologies for 6G and 5G-Advanced radio systems. He received the B.Eng. and M.Eng. (Hons.) degrees in Electrical and Electronic Engineering from Ritsumeikan University, Kyoto, Japan, in 2017 and 2019, respectively, and the Ph.D. degree in Electrical Engineering with special distinction (summa cum laude) from Jacobs University Bremen, Germany, in 2022. From 2019 to 2022, he was a Research Associate at Jacobs University Bremen on a collaborative project with Continental AG; he also held guest and visiting researcher appointments at the University of Electro-Communications, Japan, and the University of Toronto, Canada. His research interests include wireless communications, MIMO systems, signal processing, machine learning, and optimization. His honors include the Ericsson Key Contributor Award (2025, 2026) and the Yoshida Scholarship Foundation Doctoral Fellowship.
\end{IEEEbiography}

\begin{IEEEbiography}[{\includegraphics[width=1in,height=1.25in,clip,keepaspectratio]{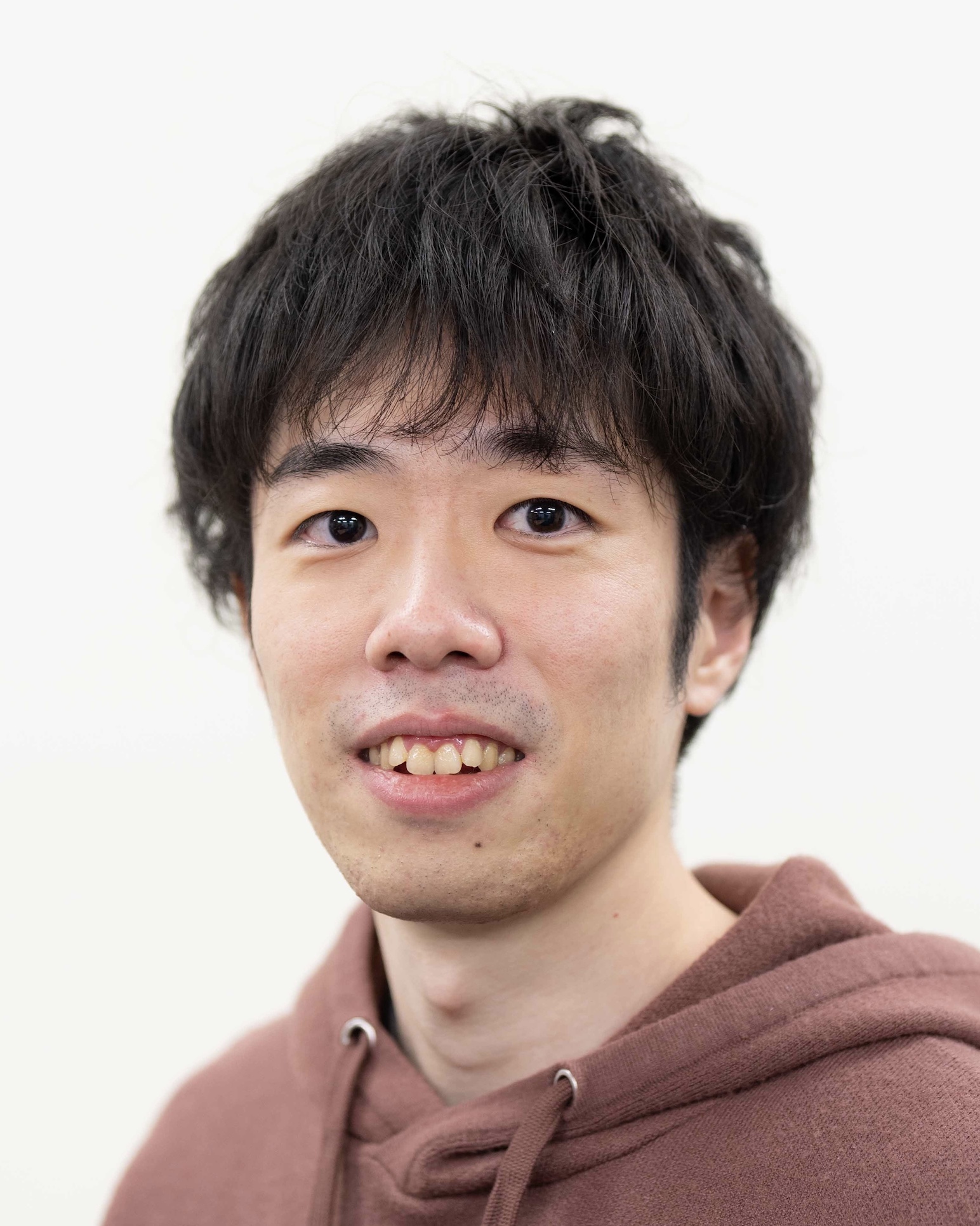}}]{Yuto Hama} 
~received the B.E., M.E., and Ph.D. degrees in information and communication engineering from Yokohama National University, Yokohama, Japan, in 2016, 2018, and 2022, respectively. From 2018 to 2021, he was a Researcher with 6G Laboratories, NTT DOCOMO, Inc., Japan. From 2021 to 2023, he was a Research Fellow of Japan Society for the Promotion of Science (JSPS). From 2022 to 2024, he was a Post-Doctoral Research Fellow at Yokohama National University. Since 2024, he has been an Experienced Researcher with the Ericsson Radio S\&R Research Laboratory, Yokohama, Japan. His research interests include fundamental physical layer design for wireless communications.
\end{IEEEbiography}

\begin{IEEEbiography}[{\includegraphics[width=1in,height=1.25in,clip,keepaspectratio]{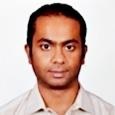}}]{Chandan Pradhan}~received the B.Tech. degree from IIIT Bhubaneswar, India, in 2013, the M.S. degree from IIIT Hyderabad, India, in 2016, and the Ph.D. degree from The University of Sydney, Sydney, NSW, Australia, in 2021. He worked as a Research Engineer at CEWiT, India, in 2016. He is currently working as a Senior Researcher at Ericsson Research, Sweden. His research interests include beamforming techniques in MIMO systems and the application of deep learning in wireless networks.
\end{IEEEbiography}

\begin{IEEEbiography}[{\includegraphics[width=1in,height=1.25in,clip,keepaspectratio]{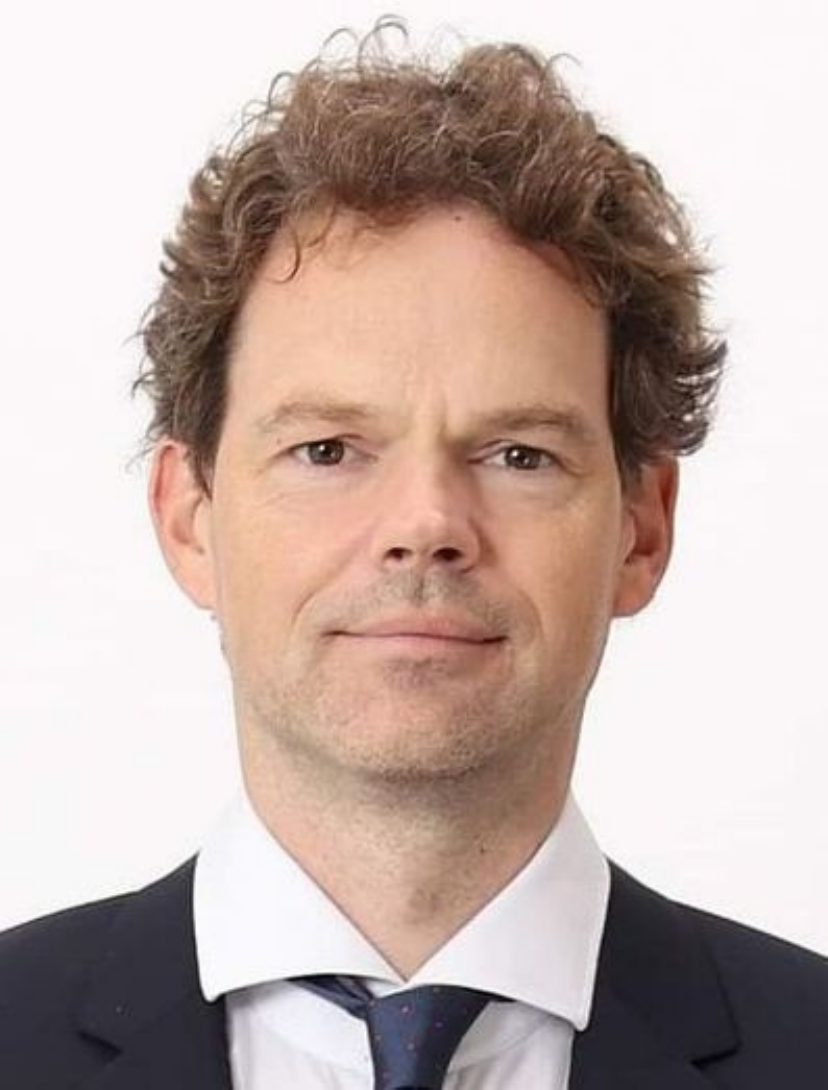}}]{Szabolcs Malomsoky}~is the head of Ericsson Research Japan. Earlier he was leading research units in Hungary, Sweden and the US. He received a Ph.D. degree from the Budapest University of Technology and Economics in 2003. Szabolcs worked with strategy setting and technical leadership in research areas including wireless technologies, network analytics, cloud computing, network management and programmable networks.
\end{IEEEbiography}

\begin{IEEEbiography}[{\includegraphics[width=1in,height=1.5in,clip,keepaspectratio]{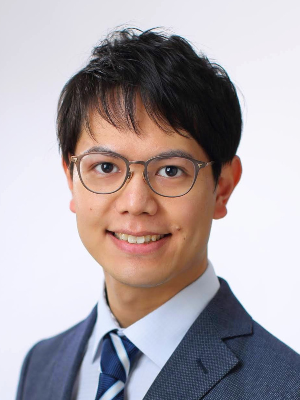}}]{Naoki~Ishikawa}
~(S'13--M'17--SM'22) is an Associate Professor with the Faculty of Engineering, Yokohama National University, Kanagawa, Japan. He received the B.E., M.E., and Ph.D. degrees from the Tokyo University of Agriculture and Technology, Tokyo, Japan, in 2014, 2015, and 2017, respectively. In 2015, he was an Academic Visitor with the School of Electronics and Computer Science, University of Southampton, U.K. From 2016 to 2017, he was a Research Fellow of the Japan Society for the Promotion of Science. From 2017 to 2020, he was an Assistant Professor in the Graduate School of Information Sciences, Hiroshima City University, Japan. In 2025, he was a Research Scholar with the Department of Electrical and Computer Engineering, University of California, San Diego, U.S. He was certified as an Exemplary Reviewer of \textsc{IEEE Transactions on Communications} in 2017 and 2021. He has served as Associate Editor of \textsc{IEEE Transactions on Vehicular Technology} since 2025. His research interests include quantum algorithms and wireless communications.
\end{IEEEbiography}

\end{document}